\documentclass[a4paper,11pt]{article}
\usepackage{pos}
\usepackage{epsfig}

\title{
\vspace*{-3.4cm}
\begin{minipage}{\textwidth}
{\normalfont\small PUBDB-2026-02625
\hspace{\fill} August 2026}\\
\end{minipage}\\[60pt]
On the status of the 4-loop splitting functions in QCD}

\ShortTitle{Status of the 4-loop splitting functions}

\author*[a]{S. Moch}
\author*[a,b]{A. Vogt$\:\!$}

\affiliation[a]{II.~Institute for Theoretical Physics, Hamburg University,
  \\ Luruper Chaussee 149, D-22761 Hamburg, Germany \\[2mm]}

\affiliation[b]{Department of Mathematical Sciences, University of Liverpool,
  \\ Liverpool L69 3BX, United Kingdom \\[2mm]}

\emailAdd{sven-olaf.moch@desy.de}
\emailAdd{Andreas.Vogt@liverpool.ac.uk}

\abstract{
We briefly discuss recent results on the four-loop splitting functions
for the scale dependence of the parton distributions of hadrons.
The three non-singlet cases are now completely known.
The results agree with almost all expectations, but exhibit some 
small-$x$ double logarithms not encountered at lower orders.
The flavour-singlet cases include, as already obvious from the present 
partial results, analytical structures not seen in the splitting 
functions up to now.
The numerical effects (and their uncertainties, in the singlet cases) 
of the four-loop contributions to the scale derivatives amount to 
about 1\% or less down to momentum fractions $x$ as low as about 
$10^{-4}$ at a standard benchmark point with $\als = 0.2$, with the 
exception of the total valence quark distribution.
}

\FullConference{Loops and Legs in Quantum Field Theory (LL2026)\\
12-17, April, 2026, Bayreuth, Germany\\}

\newcommand{\beq}{\begin{equation}}
\newcommand{\eeq}{\end{equation}}
\newcommand{\bea}{\begin{eqnarray}}
\newcommand{\eea}{\end{eqnarray}}
\newcommand{\nn}{\nonumber}

\newcommand{\hspp}{{\hspace{4mm}}}
\newcommand{\hspn}{{\hspace{-4mm}}}

\newcommand{\lsim}{\raisebox{-0.7mm}{$\:\stackrel{<}{{\scriptstyle
 \sim}}\: $} }

\def\frct#1#2{\mbox{\small{$\displaystyle\frac{#1}{#2}$}}}

\newcommand{\ra}{\rightarrow}

\newcommand{\binomial}[2]{{#1\choose #2}}

\def\nc{{n_{c}}}

\def\nf{{n^{}_{\! f}}}
\def\nfz{{n^{\,0}_{\! f}}}
\def\nfo{{n^{\,1}_{\! f}}}
\def\nfs{{n^{\,2}_{\! f}}}
\def\nft{{n^{\,3}_{\! f}}}

\def\cf{{C^{}_F}}
\def\cfs{{C^{\, 2}_F}}
\def\cft{{C^{\, 3}_F}}

\def\ca{{C^{}_A}}
\def\cas{{C^{\, 2}_A}}
\def\cat{{C^{\, 3}_A}}

\def\cfa{{C^{}_{F\!A}}}
\def\cfas{{C^{\, 2}_{F\!A}}}
\def\cfat{{C^{\, 3}_{F\!A}}}

\def\dfRAnc{{ {d_{F}^{\,abcd}d_{A}^{\,abcd}\! / n_c} }}

\def\dabcnc{{\frac{d^{abc}d_{abc}}{n_c}}}
\def\dfFAnc{{\frac{d_F^{\,abcd}d_A^{\,abcd}}{n_c}}}
\def\dfFFnc{{\frac{d_F^{\,abcd}d_F^{\,abcd}}{n_c}}}

\def\DFAnc{{D_{F\!A}}}
\def\DFFnc{{D_{F\!F}}}

\newcommand{\ep}{\varepsilon}
\newcommand{\MSb}{$\overline{\mbox{MS}}$}

\newcommand{\als}{\alpha_{\rm s}}
\newcommand{\ars}{a_{\rm s}}
\def\as(#1){{\alpha_{\rm s}^{\:#1}}}
\def\ar(#1){{a_{\rm s}^{\:#1}}}

\def\b#1{{\beta_{#1}}}
\def\B(#1,#2){{\beta_{#1}^{\,#2}}}

\def\z#1{{\zeta_{#1}^{}}}
\def\zz(#1,#2){{\zeta_{#1}^{\,#2}}}

\def\S(#1){{{S}_{#1}}}
\def\BS(#1){{{\mathbb S}_{#1}}}

\begin{document}
\maketitle

\section{Introduction}

\vspace*{-1mm}
Quark and gluon parton distribution functions (PDFs) are indispensable 
ingredients for all analyses of hard-scattering observables in 
collisions involving initial-state hadrons.
The PDFs are not calculable in perturbative QCD, unlike their 
scale dependence (evolution)
\beq
\label{eq:evol}
  \frac{\partial}{\partial \ln \mu^2} \, f_i^{}(x,\mu^2) \:=\: \sum_{k} 
 \left[ {P^{}_{\! ik}(\als(\mu^2))} \otimes f_k^{}(\mu^2) 
 \right]\!(x)
\; .
\eeq
Here $\otimes$ abbreviates the Mellin convolution in the momentum 
fraction $x$, and $\mu$ is the factorization and renormalization scale, 
usually defined in the \MSb\ scheme. 
This set of $(2\nf\!+\!1) \times (2\nf\!+\!1)$ integro-differential
equations, where $\nf$ is the number of light quark flavours, 
can be decomposed into $(2\nf - 1)$ scalar (non-singlet) equations for 
combinations of (anti-)$\,$quark distributions decoupled from each other
and the gluon distribution $g(x,\mu^2)$ and the $2 \times 2$ 
singlet-quark gluon ($q_{\rm s}^{}, g)$ system.  

\vspace*{0.5mm}
The corresponding splitting functions are 
\beq
\label{eq:Pdecomp}
  P_{\!\rm ns}^{\:\!\pm} \;,\;\;
  P_{\!\rm v}^{} \,=\, P_{\!\rm ns}^{\:\!-} + P_{\!\rm ns}^{\,\rm s} 
  \quad\mbox{and}\quad
  {\bf {P}} \,=\, \left( 
      \begin{array}{cc} \! P_{\!\rm qq} \! & \! P_{\!\rm qg} \!\! \\[-1mm]
                        \! P_{\!\rm gq} \! & \! P_{\!\rm gg} \!\! \end{array}
              \right)
  \;\mbox{ with }\;
  P_{\!\rm qq} \,=\, P_{\!\rm ns}^{\:\!+} + P_{\!\rm ps}
\eeq
for flavour differences of quark- antiquarks sums (+) and differences 
(-), the total valence distribution (v) and the ($q_{\rm s}, g)$ system. 
Their expansion in powers of the strong coupling $\als$ can be written as
\beq
\label{eq:Pexp}
  P_{\! a}(x,\als) \,=\, \sum_{m\,=\,0}  P_{\! a}^{(m)}(x) \: \ar(m+1)
\quad \mbox{with} \quad
   \ars \,=\, \als(\mu^2) / (4 \pi)
\; .
\eeq

\vspace*{0.5mm}
The even- or odd-$N$ moments of the splitting functions are related to 
the anomalous dimensions $\gamma_{a}^{(n)}(N)$ of twist-two operators in 
the light-cone expansion,
\beq
\label{eq:PvsGam}
  \gamma_{a}^{(n)}(N)
  \: = \: - P_{\! a}^{(n)}(N)
  \: = \: - \int_0^1 \!dx\, x^{N-1} P_{\! a}^{(n)}(x)
\; ,
\eeq
where the relative sign is a standard convention.
In general, eq.~(\ref{eq:PvsGam}) holds at even $N$ for $a = +$ and the
singlet ($s$) cases and at odd $N$ for $a = -,{\rm v}$. 
The remaining moments of the splitting functions are not accessible via the 
operator-product expansion (OPE) or related approaches; they involve higher 
values of the Riemann $\zeta$-function than the above `natural' moments.

\vspace*{0.5mm}
Together with the $n^{\rm th}$-order corrections to the lowest-order
partonic cross sections, the sum to $m=n$ in eq.~(\ref{eq:Pexp}) form the
(next-to-)$^n$-leading order (N$^n$LO) approximation of perturbative QCD.
The 3-loop ($n=2$) splitting functions, required for fully consistent
N$^2$LO analyses which are now the standard for many processes, were 
computed more than 20 years ago \cite{Moch:2004pa,Vogt:2004mw}.

\vspace*{0.5mm}
N$^3$LO calculations are desirable or required for very high accuracy, 
such as the determination of PDFs and the strong coupling $\als$ from 
deep-inelastic scattering (DIS) \cite{Vermaseren:2005qc,Moch:2008fj,%
Currie:2018fgr,Gehrmann:2018odt}, or for processes with large 
higher-order corrections, such as Higgs-boson production in proton-proton
collisions \cite{Anastasiou:2015vya,Mistlberger:2018etf, Chen:2021isd}.
First corresponding 4-loop contributions to all splitting functions have
been presented at a previous Loops \& Legs conference, 10 years ago 
\cite{Ruijl:2016pkm}. Since then a considerable amount of research has 
been devoted to these functions 
\cite{Davies:2016jie,Moch:2017uml,Davies:2017hyl,Moch:2018wjh,Vogt:2018miu,
Moch:2021qrk,Falcioni:2022fdm,Falcioni:2023luc,Falcioni:2023vqq,
Gehrmann:2023cqm,Gehrmann:2023iah,Falcioni:2023tzp,Moch:2023tdj,
Falcioni:2024xyt,Falcioni:2024xav,Gehrmann:2024ggw,Falcioni:2024qpd,
Kniehl:2025jfs,Kniehl:2025ttz,Falcioni:2025hfz,Kniehl:2026eij},
for now culminating in the completion of all three non-singlet splitting
functions $P_{\rm ns}^{(3)a}(x)$, $a = \pm,s$, in eqs.~(\ref{eq:Pdecomp}) 
and (\ref{eq:Pexp}) in ref.~\cite{Gehrmann:2026qbl}. 
In~this contribution, we briefly address a few aspects of these non-singlet
results, for a more detailed discussion see ref.~\cite{Moch:2026qsw}, 
as well as the present status in the flavour-singlet sector.

\section{The non-singlet cases}
\setcounter{equation}{0}

\subsection{General properties}

Until a few months ago, only the fermionic ($\:\!\nf$) contributions to 
$P_{\rm ns}^{(3)\pm}(x)$ were fully known in QCD and its gauge-group 
generalizations \cite{Davies:2016jie,Kniehl:2025ttz}.
Before ref.~\cite{Gehrmann:2026qbl}, the exact expressions for the main 
($\nfz$) contribution had been known only in the limit of a large number 
of colours $n_c$ \cite{Moch:2017uml}. 

\vspace*{0.5mm}
The corresponding anomalous dimensions to now (at least) four loops can 
be cast in the form
\beq
\label{eq:gamStr}
  \gamma_{\rm ns}^{(\ell)\pm}(N) \;=\;
  \sum_{n\,=\,0}^{2\ell-1} \;
  \sum_{a\,=\,0,1}
  \sum_{k\,=\,0}^{2\ell+1-n\,} \;
  \sum_{\vec{w},\,w\,=\,0}^{2\ell+1-n-k}
  c_{akw}^{(\ell,n)} \:
  \zeta_{n}^{} \: \frac{1}{(N\!+\!a)^{k}} \: S_{\vec{w}}(N) 
\eeq
in terms of harmonic sums recursively defined~by \cite{Vermaseren:1998uu}
\beq
\label{eq:Hsums}
  S_{\pm m}(N) \:=\: \sum_{n\,=\,1}^{N}\: (\pm 1)^n \, \frac{1}{n^{\:\!m}}
 \; , \quad
  S_{\pm m_1^{},\,m_2^{},\,\ldots,\,m_d^{}}(N) \:=\: \sum_{n\,=\,1}^{N}\:
  (\pm 1)^{n} \, \frac{1}{n^{\:\!m_1^{}}}\: S_{m_2^{},\,\ldots,\,m_d^{}}(n)
\:\: .
\eeq
The weight $w$ of a sum is given by the sum of the absolute values of 
its indices~$m_i$.  The sum over $\vec{w}$ in eq.~(\ref{eq:gamStr}) 
excludes any sums with an index -1 (see below). 
$\zeta_{n}^{}$ are values of Riemann's $\zeta$-function; the non-$\zeta$
terms are included in eq.~(\ref{eq:gamStr}) via $\zeta_{0}^{} \equiv 1$. 
The sum over $n$ excludes $n=1,2$.
The case of $\gamma_{\rm ns}^{\,\rm s}(N)$ is analogous, but the sum 
over the denominator shifts $a$ also includes $a=-1$ and $a=2$.

\vspace*{0.5mm}
The conformal symmetry of QCD for $D = 4 -2\:\!\ep$ dimensions at a 
particular value of the coupling constant implies the relation
\cite{Dokshitzer:2005bf,Dokshitzer:2006nm,Basso:2006nk}, 
see also ref.~\cite{Strohmaier:2018tjo},
\beq
\label{eq:gamma-u}
  \gamma_{\rm ns}^{\,a}(N)  \:=\: \gamma_{\rm u}^{\,a}
  \left( N + \sigma \,\gamma_{\rm ns}^{\,a}(N)-\beta(\ars)/\ars ) \right)
\; ,
\eeq
with $\beta(\ars) = - \beta_0\,\ar(2) - \beta_1\,\ar(3) - \dots$, 
and $\,\sigma = -1\,$ for the present initial-state (space-like)
splitting functions and $\,\sigma = 1\,$ for their final-state (time-like)
fragmentation counterparts.  Eq.~(\ref{eq:gamma-u}) leads to
\bea
\label{eq:gu-exp}
  \gamma_{\rm u}^{} \:=\:
         \sum_{n\,=\,0}\: \ar(n+1)\, \gamma_{\rm u}^{(n)}
   &\!=\!& \:\:
         \ars \,\* \gamma_0^{}
\nn \\[-3mm] & & \mbox{\hspn}
   \,+\, \ar(2) \* \left( \,
         \overline{\gamma}_1^{}
      \,-\, \b0 \,\* d_N^{} \* \gamma_0^{}
       \right)
\nn \\ & & \mbox{\hspn}
   \,+\, \ar(3) \* \Big( \,
         \overline{\gamma}_2^{}
      - \frct{1}{6}\, \* d_N^{\,2\,} \* \gamma_0^{\:3}
      - \b1 \,\* d_N^{} \* \gamma_0^{}
      - \b0 \,\* d_N^{} \* \overline{\gamma}_1^{}
      + \frct{1}{2}\, \* \B(0,2) \,\* d_N^{\,2\,} \* \gamma_0^{}
      \Big)
\nn \\ & & \mbox{\hspn}
   \,+\, \ar(4) \* \Big( \,
         \overline{\gamma}_3^{}
      - \frct{1}{2}\, \* d_N^{\,2\,} \*
         ( \gamma_0^{\:2}\,\overline{\gamma}_1^{} )
      - \b2\,\* d_N^{} \* \gamma_0^{}
      - \b1\,\* d_N^{} \* \overline{\gamma}_1^{}
      - \b0\,\* d_N^{} \* \overline{\gamma}_2^{}
\nn \\ & & \mbox{\hspp}
      + \frct{1}{6}\, \* \b0 \,\* d_N^{\,3\,} \* \gamma_0^{\:3}
      + \b0\,\* \b1\,\* d_N^{\,2\,} \* \gamma_0^{}
      + \frct{1}{2}\, \* \B(0,2) \,\* d_N^{\,2\:} \* \overline{\gamma}_1^{}
      - \frct{1}{6}\, \* \B(0,3) \,\* d_N^{\,3\:} \* \gamma_0^{}
      \Big)
\nn \\[1mm] & & \mbox{\hspn}
  \,+\, {\cal O}(\ar(5))
\; ,
\eea
where all arguments $N$ have been suppressed for brevity.
$\overline{\gamma}_n^{}$ denotes the average of the $\sigma = -1$ and 
$\sigma = 1$ expansion coefficients. 
Since the difference between the time-like and space-like non-singlet
splitting functions is fixed by lower-order information, 
see refs.~\cite{Mitov:2006ic,Moch:2017uml} and references therein,
eq.~(\ref{eq:gu-exp}) can be written purely in terms of the present 
$\sigma = -1$ parton-evolution splitting functions.

\vspace*{0.5mm}
We have evaluated the derivatives $d_N^{\,m} \equiv d^{\,m}/dN^{\,m}$ in 
eq.~(\ref{eq:gu-exp}) via inverse Mellin transforms to $x$-space, where 
they correspond to products with $\ln^{\,m\!}x$, and subsequent Mellin 
transforms of these products in {\sc Form} 
\cite{Vermaseren:2000nd,Kuipers:2012rf,Ruijl:2017dtg}
using packages first provided with ref.~\cite{Vermaseren:2000nd}.

A salient point of eq.~(\ref{eq:gu-exp}) is that the universal evolution 
kernels $\gamma_{\rm u}^{(n)a}(N)$ are reciprocity respecting (RR), i.e., 
their Mellin inverses fulfil $P_{\rm u}^{(n)a}(x) = - x P_{\rm u}^{(n)a}(1/x)$.
For the `unsummed' denominators analogous to eq.~(\ref{eq:gamStr}) this 
implies that they can occur only in the combinations 
\beq
\label{eq:RRdenoms}
  \eta \:=\: \frct{1}{N} - \frct{1}{N\!+\!1}
\; , \;\;
  \nu  \:=\: \frct{1}{N\!-\!1} - \frct{1}{N\!+\!2}
\; ,
\eeq
respectively corresponding to $\,1-x$ and $\,x^{-1} - x^2$, which are 
invariant under $N \ra -N\!-\!1$. 
The harmonic sums can enter only in suitable linear combinations
which can be expressed in~terms~of
\beq
\label{eq:Bsums}
  {\mathbb S}_{m_1,\dots,m_d^{}}(N) \:=\:
  (-1)^N \sum_{k\,=\,1}^{N}(-1)^k \binomial{N}{k}\binomial{N+k}{k}\,
  S_{m_1^{},...,m_d^{}}(k)
\; ,
\eeq
where all indices are positive.
These binomial sums, already considered at $w \leq 4$ in ref.~%
\cite{Vermaseren:1998uu}, provide a basis of all RR linear combinations 
of the harmonic sums (\ref{eq:Hsums}).

\vspace*{0.5mm}
Other bases are possible which exploit the fact that products of RR
sums are again RR. Since they do not lead to (significantly) shorter 
expressions at high weights, it seems easiest to use eq.~(\ref{eq:Bsums}).
A set of equations for converting RR functions given in terms of standard 
harmonic sums up to weight 7 to binomial sums can be found in app.~B of 
ref.~\cite{Moch:2026qsw}.

\vspace*{0.5mm}
The fact the $\gamma_{\rm u}^{(n)a}(N)$ in eq.~(\ref{eq:gu-exp}) can be 
expressed in terms of the quantities (\ref{eq:RRdenoms}) and 
(\ref{eq:Bsums}) has direct consequences on the large-$N$ ($x \!\ra\! 1$) 
limit. This has been discussed in refs.~\cite{Gehrmann:2026qbl,Moch:2026qsw}, 
we do not need to repeat it here.
It may be worthwhile to note, however, that this RR structure provides a 
partial explanation for the observation that sums with index -1 do not occur 
in any splitting functions in massless perturbative QCD (we do not recall 
any derivation of this in the literature): there are no RR combinations of
harmonic sums that involve any sums with index -1. Hence such sums can
enter neither $\gamma_{\rm u}^{(n)a}$ nor, after inverting 
eq.~(\ref{eq:gu-exp}), the non-singlet anomalous dimensions in the \MSb\ 
scheme.
The analogous statement in the singlet sector is, at this point, restricted 
to the trace and determinant of the matrix in eq.~(\ref{eq:Pdecomp}), see 
ref.~\cite{Kniehl:2026eij}.

\subsection{Small-$x$ expansions}
\allowdisplaybreaks[2]

\vspace*{-1mm}
The most intriguing aspect of the new results for
$\gamma_{\rm ns}^{(n)a}$ of ref.~\cite{Gehrmann:2026qbl} is a deviation
from (our) previous expectations \cite{Vogt:2012gb,Davies:2022ofz}
for the small-$x$ expansion. We have discussed this issue already in 
ref.~\cite{Moch:2026qsw}; here we address it in a different (but related, 
of course) manner. 

\vspace*{0.5mm}
Transforming the coefficients $\gamma_u^{(n)a}$ in eq.~(\ref{eq:gu-exp}),
which we have provided in an ancillary file to ref.~\cite{Moch:2026qsw}, 
to $x$-space and expanding in powers of $x$ and $\ln x$,
\beq
\label{eq:Psmallx}
  P_{\rm u}^{(n)a}(x) \:=\: \sum_{p\,=\,0} \: \sum_{\ell=\,0}^{2n} \,
     x^p \ln^{2n-\ell} \! x \;{\mathbb P}_{n,a}^{\,(p,\ell)}
\; ,
\eeq
one arrives at the double-log (DL, $\ell < n$) coefficients of 
$P_{\rm u}^{\,+}(x)$ for the first four values of $p$:
\bea
\label{eq:Pp1smallx}
  {\mathbb P}_{1,+}^{(0,0)} &\!=\! & 0
\nn \\
  {\mathbb P}_{1,+}^{(1,0)} &\!=\! & \mbox{}
  - 8\,\*\cf\*\cfa
\nn \\
  {\mathbb P}_{1,+}^{(2,0)} &\!=\! & 0
\nn \\
  {\mathbb P}_{1,+}^{(3,0)} &\!=\! & \mbox{}
  - 16\,\*\cf\*\cfa
\; ,
\eea
\bea
\label{eq:Pp2smallx}
  {\mathbb P}_{2,+}^{(0,0)} &\!=\! & {\mathbb P}_{2,+}^{(0,1)} \:=\: 0
\nn \\
  {\mathbb P}_{2,+}^{(1,0)} &\!=\! & \mbox{}
  - 4\,\*\cf\*\cfas
  + \frct{16}{3}\,\*\cfs\*\cfa
\nn \\
  {\mathbb P}_{2,+}^{(1,1)} &\!=\! & \mbox{}
  - \frct{368}{9}\,\*\cf\*\cfas
  - \frct{16}{9}\,\*\cf\*\nf\*\cfa
  + \frct{440}{9}\,\*\cfs\*\cfa
\nn \\[2mm]
  {\mathbb P}_{2,+}^{(2,0)} &\!=\! & {\mathbb P}_{2,+}^{(2,1)} \:=\: 0
\nn \\
  {\mathbb P}_{2,+}^{(3,0)} &\!=\! & \mbox{}
  - \frct{32}{3}\,\*\cf\*\cfas
  + 16\,\*\cfs\*\cfa
\nn \\
  {\mathbb P}_{2,+}^{(3,1)} &\!=\! & \mbox{}
  - \frct{1696}{9}\,\*\cf\*\cfas
  - \frct{32}{9}\,\*\cf\*\nf\*\cfa
  + 128\,\*\cfs\*\cfa
\; , \\[-7mm] \nn
\eea
%
%
\bea
\label{eq:Pp3smallx}
  {\mathbb P}_{3,+}^{(0,0)} &\!=\! & {\mathbb P}_{3,+}^{(0,1)} \:=\: 0
\nn \\
  {\mathbb P}_{3,+}^{(0,2)} &\!=\! & \mbox{}
  \zeta_2 \* \left(
  - 48\,\*\DFAnc
  - 288\,\*\cf\*\cfat
  + 384\,\*\cfs\*\cfas
  - 112\,\*\cft\*\cfa
  \right)
\nn \\[2mm]
  {\mathbb P}_{3,+}^{(1,0)} &\!=\! & \mbox{}
  - \frct{32}{15}\,\*\cf\*\cfat
  + \frct{152}{45}\,\*\cfs\*\cfas
  - \frct{92}{45}\,\*\cft\*\cfa
\nn \\
  {\mathbb P}_{3,+}^{(1,1)} &\!=\! & \mbox{}
  - \frct{376}{15}\,\*\cf\*\cfat
  - \frct{8}{15}\,\*\cf\*\nf\*\cfas
  + \frct{2252}{45}\,\*\cfs\*\cfas
  + \frct{8}{9}\,\*\cfs\*\nf\*\cfa
  - \frct{1448}{45}\,\*\cft\*\cfa
\nn \\
  {\mathbb P}_{3,+}^{(1,2)} &\!=\! & \mbox{}
  - \frct{32}{3}\,\*\DFAnc
  + \frct{2992}{27}\,\*\cf\*\cfat
  - \frct{104}{27}\,\*\cf\*\nf\*\cfas
  - \frct{8}{27}\,\*\cf\*\nfs\*\cfa
  - \frct{2984}{27}\,\*\cfs\*\cfas
\nn \\ &&  \mbox{}
  + \frct{128}{27}\,\*\cfs\*\nf\*\cfa
  + \frct{2836}{27}\,\*\cft\*\cfa
  + \zeta_2 \* \left(
    16\*\DFAnc
  + \frct{704}{3}\,\* \big( \cf\*\cfat - \cfs\*\cfas \big)
  + 80\*\cft\*\cfa
  \right)
\nn \\[2mm]
  {\mathbb P}_{3,+}^{(2,0)} &\!=\! & {\mathbb P}_{3,+}^{(2,1)} \:=\: 0
\nn \\
  {\mathbb P}_{3,+}^{(2,2)} &\!=\! & \mbox{}
  \zeta_2 \* \left(
  - 128\,\*\DFAnc
  - 704\,\*\cf\*\cfat
  + 928\,\*\cfs\*\cfas
  - 288\,\*\cft\*\cfa
  \right)
\nn \\[2mm]
  {\mathbb P}_{3,+}^{(3,0)} &\!=\! & \mbox{}
  - \frct{64}{9}\,\*\cf\*\cfat
  + \frct{32}{3}\,\*\cfs\*\cfas
  - \frct{64}{9}\,\*\cft\*\cfa
\nn \\
  {\mathbb P}_{3,+}^{(3,1)} &\!=\! & \mbox{}
  - \frct{160}{9}\,\*\DFAnc
  - \frct{4768}{45}\,\*\cf\*\cfat
  - \frct{64}{45}\,\*\cf\*\nf\*\cfas
  + \frct{10352}{45}\,\*\cfs\*\cfas
  + \frct{128}{45}\,\*\cfs\*\nf\*\cfa
\nn \\ &&  \mbox{}
  - \frct{4624}{45}\,\*\cft\*\cfa
\nn \\
  {\mathbb P}_{3,+}^{(3,2)} &\!=\! & \mbox{}
  - \frct{8260}{27}\,\*\DFAnc
  - \frct{544}{27}\,\*\nf\*\DFFnc
  + \frct{5248}{9}\,\*\cf\*\cfat
  + \frct{208}{27}\,\*\cf\*\nf\*\cfas
  - \frct{16}{27}\,\*\cf\*\nfs\*\cfa
\nn \\ &&  \mbox{}
  - \frct{2696}{9}\,\*\cfs\*\cfas
  - \frct{208}{9}\,\*\cfs\*\nf\*\cfa
  + \frct{11744}{27}\,\*\cft\*\cfa
\nn \\ &&  \mbox{}
  + \zeta_2 \* \left(
    \frct{1472}{3}\,\*\cf\*\cfat
  - \frct{1568}{3}\,\*\cfs\*\cfas
  + 224\,\*\cft\*\cfa
  \right)
\; .
\eea
Here and below we have used the abbreviations $\cfa = C_F - 1/2\,C_A$ and
\beq
  \DFFnc \,=\, \dfFFnc - \frct{1}{48}\, \cas \cf\, ,\quad
  \DFAnc \,=\, \dfFAnc - \frct{1}{24}\, \cat \cf
\eeq
for large-$\nc$ suppressed combinations of group invariants.

\vspace{0.5mm}
The coefficients of the even powers $p$ of $x$ in eqs.~(\ref{eq:Pp1smallx}),
(\ref{eq:Pp2smallx}) vanish; the odd-$p$ coefficients are non-zero 
already at the leading-log (LL, $\ell=0\,$) level, except in the 
large-$n_c$ (L$n_c$) limit of SU($\nc$).
The surprising (to us) result of ref.~\cite{Gehrmann:2026qbl} is that this 
even-$p$ pattern does not hold at four loops ($n=3$). 
Non-vanishing $\z2 = \pi^2/6$ next-to-next-to-leading log (NNLL, $\ell=2$) 
coefficients occur for all $\nfz$ colour factors; the $\ell=2$ contribution 
only vanishes in the L$n_c$ limit. 
Consequently the even-$p$ $x^p \ln^{\,4\!} x$ terms of $P_{\rm ns}^{(3)+}(x)$ 
are not fixed by lower-order results via eq.~(\ref{eq:gu-exp}), 
and the $x^0$-predictions of refs.~\cite{Vogt:2012gb,Davies:2022ofz} agree, 
for the $\pi^2$ terms, 
only at large $\nc$ with the result of~ref.~\cite{Gehrmann:2026qbl}.

\vspace{0.5mm}
The corresponding coefficients for $P_{\,\rm u}^{\,-}(x)$ in the small-$x$
expansion (\ref{eq:Psmallx}) are given by
\bea
\label{eq:Pm1smallx}
  {\mathbb P}_{1,-}^{(0,0)} &\!=\! &
  - 8\,\*\cf\*\cfa
\nn \\
  {\mathbb P}_{1,-}^{(1,0)} &\!=\! & 0
\nn \\
  {\mathbb P}_{1,-}^{(2,0)} &\!=\! &
  - 16\,\*\cf\*\cfa
\nn \\
  {\mathbb P}_{1,-}^{(3,0)} &\!=\! & 0
\; , \\[-6mm] \nn
\eea
\bea
\label{eq:Pm2smallx}
  {\mathbb P}_{2,-}^{(0,0)} &\!=\! & \mbox{}
  - 4\,\*\cf\*\cfas
  + \frct{16}{3}\,\*\cfs\*\cfa
\nn \\
  {\mathbb P}_{2,-}^{(0,1)} &\!=\! & \mbox{}
  - \frct{368}{9}\,\*\cf\*\cfas
  - \frct{16}{9}\,\*\cf\*\nf\*\cfa
  + \frct{440}{9}\,\*\cfs\*\cfa
\nn \\[2mm]
  {\mathbb P}_{2,-}^{(1,0)} &\!=\! & {\mathbb P}_{2,-}^{(1,1)} \:=\: 0
\nn \\
  {\mathbb P}_{2,-}^{(2,0)} &\!=\! & \mbox{}
  - \frct{32}{3}\,\*\cf\*\cfas
  + 16\,\*\cfs\*\cfa
\nn \\
  {\mathbb P}_{2,-}^{(2,1)} &\!=\! & \mbox{}
  - \frct{928}{9}\,\*\cf\*\cfas
  - \frct{32}{9}\,\*\cf\*\nf\*\cfa
  + \frct{640}{9}\,\*\cfs\*\cfa
\nn \\
  {\mathbb P}_{2,-}^{(3,0)} &\!=\! & {\mathbb P}_{2,-}^{(3,1)} \:=\: 0
\; , \\[-6mm] \nn
\eea
%
%
\bea
\label{eq:Pm3smallx}
  {\mathbb P}_{3,-}^{(0,0)} &\!=\! & \mbox{}
  - \frct{32}{15}\,\*\cf\*\cfat
  + \frct{152}{45}\,\*\cfs\*\cfas
  - \frct{92}{45}\,\*\cft\*\cfa
\nn \\
  {\mathbb P}_{3,-}^{(0,1)} &\!=\! & \mbox{}
  - \frct{376}{15}\,\*\cf\*\cfat
  - \frct{8}{15}\,\*\cf\*\nf\*\cfas
  + \frct{2252}{45}\,\*\cfs\*\cfas
  + \frct{8}{9}\,\*\cfs\*\nf\*\cfa
  - \frct{1448}{45}\,\*\cft\*\cfa
\nn \\
  {\mathbb P}_{3,-}^{(0,2)} &\!=\! & \mbox{}
  - \frct{272}{27}\,\*\cf\*\cfat
  + \frct{280}{27}\,\*\cf\*\nf\*\cfas
  - \frct{8}{27}\,\*\cf\*\nfs\*\cfa
  - \frct{3032}{27}\,\*\cfs\*\cfas
  - \frct{448}{27}\,\*\cfs\*\nf\*\cfa
\nn \\ &&  \mbox{}
  + \frct{1396}{27}\,\*\cft\*\cfa
  + \zeta_2 \* \left(
    16\*\DFAnc
  + \frct{704}{3}\,\*\cf\*\cfat
  - \frct{704}{3}\,\*\cfs\*\cfas
  + 80\*\cft\*\cfa
  \right)
\nn \\[2mm]
  {\mathbb P}_{3,-}^{(1,0)} &\!=\! & {\mathbb P}_{3,-}^{(1,1)} \:=\: 0
\nn \\
  {\mathbb P}_{3,-}^{(1,2)} &\!=\! & \mbox{}
  \zeta_2 \* \left(
    - 48\,\*\DFAnc
    - 288\,\*\cf\*\cfat
    + 384\,\*\cfs\*\cfas
    - 112\,\*\cft\*\cfa
  \right)
\nn \\[2mm]
  {\mathbb P}_{3,-}^{(2,0)} &\!=\! & \mbox{}
  - \frct{64}{9}\,\*\cf\*\cfat
  + \frct{32}{3}\,\*\cfs\*\cfas
  - \frct{64}{9}\,\*\cft\*\cfa
\nn \\
  {\mathbb P}_{3,-}^{(2,1)} &\!=\! & \mbox{}
  - \frct{32}{5}\,\*\DFAnc
  - \frct{544}{9}\,\*\cf\*\cfat
  - \frct{64}{45}\,\*\cf\*\nf\*\cfas
  + \frct{400}{3}\,\*\cfs\*\cfas
  + \frct{128}{45}\,\*\cfs\*\nf\*\cfa
  - \frct{2704}{45}\,\*\cft\*\cfa
\nn \\
  {\mathbb P}_{3,-}^{(2,2)} &\!=\! & \mbox{}
  - \frct{292}{3}\,\*\DFAnc
  - \frct{32}{3}\,\*\nf\*\DFFnc
  + \frct{18560}{27}\,\*\cf\*\cfat
  + \frct{464}{27}\,\*\cf\*\nf\*\cfas
  - \frct{16}{27}\,\*\cf\*\nfs\*\cfa
\nn \\ &&  \mbox{}
  - \frct{27352}{27}\,\*\cfs\*\cfas
  - \frct{944}{27}\,\*\cfs\*\nf\*\cfa
  + \frct{18080}{27}\,\*\cft\*\cfa
\nn \\ &&  \mbox{}
  + \zeta_2 \* \left(
      \frct{1472}{3}\,\*\cf\*\cfat
    - \frct{1568}{3}\,\*\cfs\*\cfas
    + 224\*\cft\*\cfa
  \right)
\nn \\[2mm]
  {\mathbb P}_{3,-}^{(3,0)} &\!=\! & {\mathbb P}_{3,-}^{(3,1)} \:=\: 0
\nn \\
  {\mathbb P}_{3,-}^{(3,2)} &\!=\! & \mbox{}
  \zeta_2 \* \left(
    - 128\,\*\DFAnc
    - 704\,\*\cf\*\cfat
    + 928\,\*\cfs\*\cfas
    - 288\,\*\cft\*\cfa
    \right)
\; .
\eea
The structure is the same as in eqs.~(\ref{eq:Pp1smallx}) -- 
(\ref{eq:Pp3smallx}), except that the roles of the even and odd powers 
of $p$ are interchanged.  The pattern for $P_{\rm u}^{\,\rm s}(x)$, 
also based on odd-$N$ moments in the OPE 
(unlike $P_{\rm u}^{\,+}(x)$, see eq.~(\ref{eq:PvsGam}) above) is 
-- up to some additional zeros --  
as in eqs.~(\ref{eq:Pm1smallx}) -- (\ref{eq:Pm3smallx}) with
\bea
\label{eq:Ps2smallx}
  {\mathbb P}_{2,\rm s}^{(0,0)} &\!=\! &
  \frct{16}{3}\,\*\nf\*\dabcnc
\nn \\
  {\mathbb P}_{2,\rm s}^{(0,1)} &\!=\! &
  - \frct{32}{3}\,\*\nf\*\dabcnc
\nn \\
  {\mathbb P}_{2,\rm s}^{(1,0)} &\!=\! & {\mathbb P}_{2,\rm s}^{(1,1)} \:=\: 0
\nn \\[1mm]
  {\mathbb P}_{2,\rm s}^{(2,0)} &\!=\! & 0
\nn \\
  {\mathbb P}_{2,\rm s}^{(2,1)} &\!=\! &
  - \frct{256}{9}\,\*\nf\*\dabcnc
\nn \\
  {\mathbb P}_{2,\rm s}^{(3,0)} &\!=\! & {\mathbb P}_{2,\rm s}^{(3,1)} \:=\: 0
\; , \\[-6mm] \nn
\eea
%
%
\bea
\label{eq:Ps3smallx}
  {\mathbb P}_{3,\rm s}^{(0,0)} &\!=\! &
  - \frct{16}{5}\,\*\ca\*\nf\*\dabcnc
\nn \\
  {\mathbb P}_{3,\rm s}^{(0,1)} &\!=\! &
  \left(
  \frct{32}{15}\,\*\nfs
  - \frct{208}{15}\,\*\ca\*\nf
  + \frct{32}{15}\,\*\cf\*\nf \right) \*\dabcnc
\nn \\
  {\mathbb P}_{3,\rm s}^{(0,2)} &\!=\! &
  \left(
  - \frct{256}{9}\,\*\nfs
  - \frct{1760}{9}\,\*\ca\*\nf
  + \frct{112}{3}\,\*\cf\*\nf 
  + \zeta_2 \* \big(
    144\,\*\ca\*\nf
  - \frct{32}{3}\,\*\cf\*\nf
  \big) \right) \*\dabcnc
\nn \\
  {\mathbb P}_{3,\rm s}^{(1,0)} &\!=\! & {\mathbb P}_{3,\rm s}^{(1,1)} \:=\: 0
\nn \\
  {\mathbb P}_{3,\rm s}^{(1,2)} &\!=\! &
  \zeta_2 \* \left(
      48\,\*\ca\*\nf
    - 32\,\*\cf\*\nf
  \right) \*\dabcnc
\nn \\
  {\mathbb P}_{3,\rm s}^{(2,0)} &\!=\! & 0
\nn \\
  {\mathbb P}_{3,\rm s}^{(2,1)} &\!=\! &
  \frct{256}{15}\,\*\ca\*\nf\*\dabcnc
\nn \\
  {\mathbb P}_{3,\rm s}^{(2,2)} &\!=\! &
  \left(
  - \frct{128}{9}\,\*\nfs
  - \frct{3520}{27}\,\*\ca\*\nf
  + \frct{832}{27}\,\*\cf\*\nf \right) \*\dabcnc
\nn \\
  {\mathbb P}_{3,\rm s}^{(3,0)} &\!=\! & {\mathbb P}_{3,\rm s}^{(3,1)} \:=\: 0
\nn \\[1mm]
  {\mathbb P}_{3,\rm s}^{(3,2)} &\!=\! & 0
\; .
\eea
 
The $\z2$ terms that fall outside the expected structure are
${\mathbb P}_{3,+}^{(0,2)}$ and ${\mathbb P}_{3,+}^{(2,2)}$ in 
eq.~(\ref{eq:Pp3smallx}),
${\mathbb P}_{3,-}^{(1,2)}$ and ${\mathbb P}_{3,-}^{(3,2)}$ in 
eq.~(\ref{eq:Pm3smallx}) and
${\mathbb P}_{3,\rm s}^{(1,2)}$ in eq.~(\ref{eq:Ps3smallx}).
The first four of these contribute to eq.~(3.15) of ref.~\cite{Moch:2026qsw},
which was derived not via the universal kernels, but following the approach
of ref.~\cite{Davies:2022ofz} for the $D$-dimensional structure of 
unfactorized structure functions.
As mentioned in ref.~\cite{Moch:2026qsw}, 
these additional terms may be related to the issues studied in 
refs.~\cite{Becher:2024kmk,Becher:2025igg} and references therein.

\vspace*{0.5mm}
The remaining non-vanishing N$^2$LO and N$^3$LO double-logarithmic 
coefficient in eqs.~(\ref{eq:Pp2smallx}) -- (\ref{eq:Ps3smallx}) --
which all vanish in the large-$\nc$ limit -- have 
not been obtained from lower-order information either, with the exception 
of the leading-log coefficients ${\mathbb P}_{2,-}^{(0,0)}$ and 
${\mathbb P}_{3,-}^{(0,0)}$ which were correctly predicted about 
30 years ago \cite{Blumlein:1995jp,Blumlein:1996aw} on the basis of 
ref.~\cite{Kirschner:1983di}.
Maybe the results collected above can be useful in future research 
towards improving upon this situation.

For the rest of this subsection, we focus on the small-$x$ expansion
of the (reciprocity-respecting)  $\nfz$ quartic Casimir contribution, 
which does not include $x^{\,0}$ terms enhanced by $\ln^{\,\ell\!} x$ 
without $\pi^2$, 
\beq
\label{eq:PpRAlnx}
  P_{\rm ns}^{\,(3)+}(x) \big|_{\,\dfRAnc} \:=\:
    -\,  48\,\* \z2\* \ln^{\,4\!} x
  \,-\, 256\,\* \z2\* \ln^{\,3\!} x
  \,-  \left(  2112\,\* \z2 -  768\,\* \zz(2,2) \right) \* \ln^{\,2\!} x
  \:+\: \dots
\:\; .
\eeq
The $\ln^{\,1\!} x$ term can be found in eq.~(3.15) of 
ref.~\cite{Moch:2026qsw}. 
The only combinations of $\eta$ in eq.~(\ref{eq:RRdenoms}) and binomial sums
in eq.~(\ref{eq:Bsums}) to overall weight 7 that contribute to the terms
shown in eq.~(\ref{eq:PpRAlnx}) are 
\beq
\label{eq:Nto0fcts}
  \eta^6 \,{\mathbb S}_1 \, , \;\; 
  \eta^5 \big\{ \,
    {\mathbb S}_1 ,\: 
    {\mathbb S}_{1,1} ,\:
    {\mathbb S}_2 \big\}
 \;\; \mbox{ and } \;\;
  \eta^4 \big\{ \,
    {\mathbb S}_1 ,\:
    {\mathbb S}_2 ,\:
    {\mathbb S}_{1,1,1} ,\: 
    {\mathbb S}_{1,2} ,\:
    {\mathbb S}_{2,1} ,\:
    {\mathbb S}_{3} \big\}
\; .
\eeq
Note that the even-$N$ expression $\gamma_{\rm ns}^{\,(3)+}(N)$ does not
include any terms with $\pi^2$. 
The part of this expression, completely given in eq.~(C.1) of 
ref.~\cite{Moch:2026qsw}, that leads to eq.~(\ref{eq:PpRAlnx}) is
\bea
  \frct{1}{16}\: \gamma_{\rm ns}^{\,(3)+}(N) \big|_{\,\dfRAnc} &\!\!=\! &
        \big( \,
          216\,\*\eta^4
        + 168\,\*\eta^5
        + 36\,\*\eta^6 
        \big) \,\* \BS(1)
      + \big( \,
          88\,\*\eta^4
        + 24\,\*\eta^5 
        \big) \,\* \BS(2)
\nn \\[-1mm] && \mbox{\hspn} \;
        - 20\,\*\eta^5 \,\* \BS(1,1)
        + 14\,\*\eta^4 \,\* \BS(1,2) 
        - 10\,\*\eta^4 \,\* \BS(2,1)
        + 40\,\*\eta^4 \,\* \BS(3)
+ \dots \; . \;\;
\eea
Here the coefficients in the first line are uniquely connected to the 
$\zz(2,1)$ part and the absence of $\z3$ in eq.~(\ref{eq:PpRAlnx}), those 
in the second line all contribute only to the coefficient of $\zz(2,2)$.

\subsection{Numerical results}
\allowdisplaybreaks[2]

\vspace*{-0.5mm}
During the past years, we have extended the fixed-$N$ {\sc Forcer} 
\cite{Ruijl:2017cxj} computations of $P_{\rm ns}^{(3)\pm}$ in 
ref.~\cite{Moch:2017uml} to $N=22$ for all $\nfz$ terms and to $N=24$
of higher for the various $\nfo$ contributions. The latter results were
used to obtain the all-$N$ expressions in ref.~\cite{Kniehl:2025ttz},
the former
 -- supplemented by the exact result for the four-loop cusp anomalous 
dimension $A_{\rm q}^{(4)}$ \cite{Henn:2019swt,vonManteuffel:2020vjv} -- 
to improve upon the approximations in ref.~\cite{Moch:2017uml}.
Of course, these newer approximations have now been superseded.

\begin{figure}[p]
\vspace{-3mm}
\centerline{\hspace*{-2mm}\epsfig{file=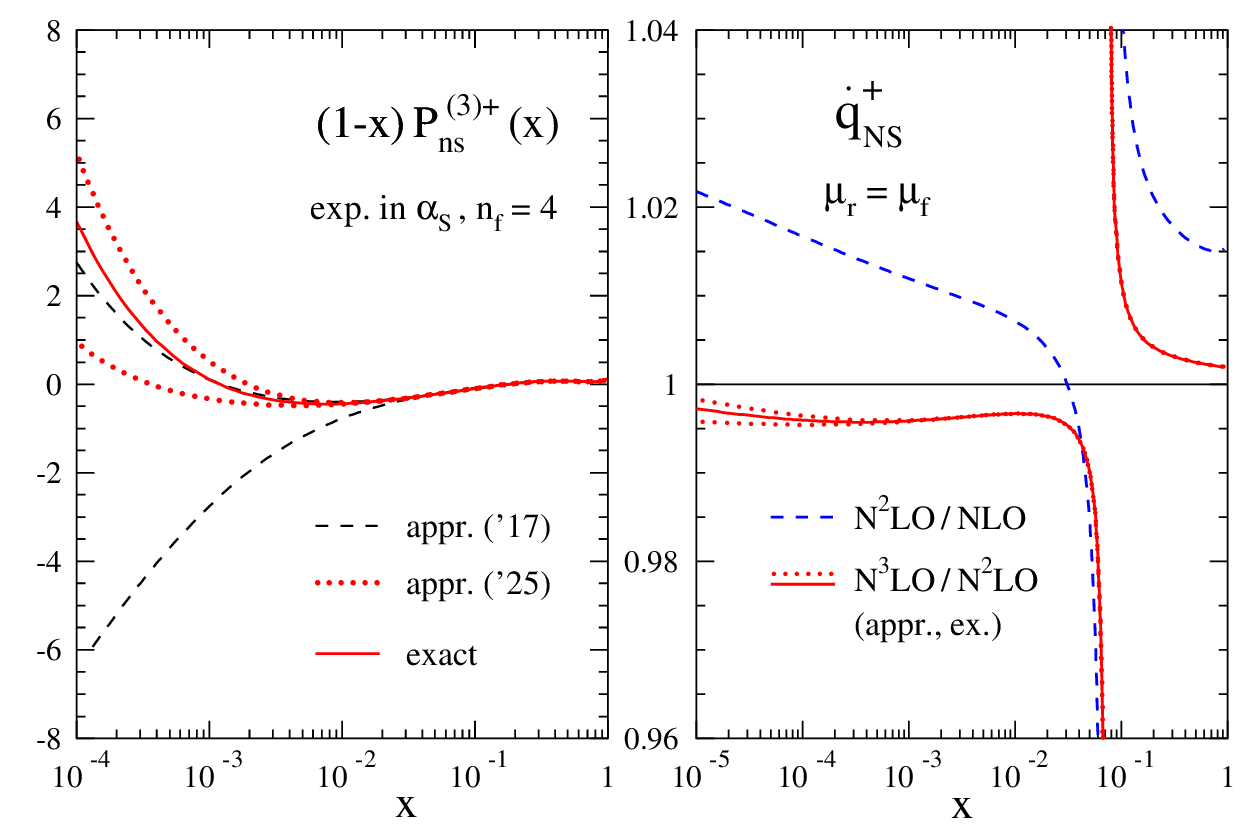,width=15.0cm,angle=0}}
\vspace{-1mm}
\caption{\label{fig:pns3p} \small
Left part: The new result \cite{Gehrmann:2026qbl} for the even-$N$ based 
splitting function $P_{\rm ns}^{(3)+}$ for four light flavours,
compared to our earlier approximations based on $N \!\leq\! 16\,$ ('17) 
\cite{Moch:2017uml} and $N \!\leq\! 22\,$ ('25) [unpubl.].
Right part: The relative N$^2$LO and N$^3$LO contributions to the
scale derivative of $q_{\rm ns}^{+}$ for the input (\ref{eq:NSinp}) and
(\ref{eq:asref}).
}
\end{figure}

\vspace*{0.5mm}
It is, however, not entirely uninteresting to compare them with the exact
results \cite{Gehrmann:2026qbl}, in particular in the case of 
$P_{\rm ns}^{(3)+}(x)$, where the exact result is slightly above the 2017
error band at $x < 10^{-3}$. 
This is done in the left panel of fig.~\ref{fig:pns3p}.
The additional information available in 2025, three more even-$N$ or 
odd-$N$ moments and the exact leading large-$x$ contribution
$A_{\rm q}^{(4)}\! / (1-x)_{+\,}$, lead to small uncertainties for 
about one additional order of magnitude in $x$ towards $x=0$.

\begin{figure}[p]
\centerline{\hspace*{-2mm}\epsfig{file=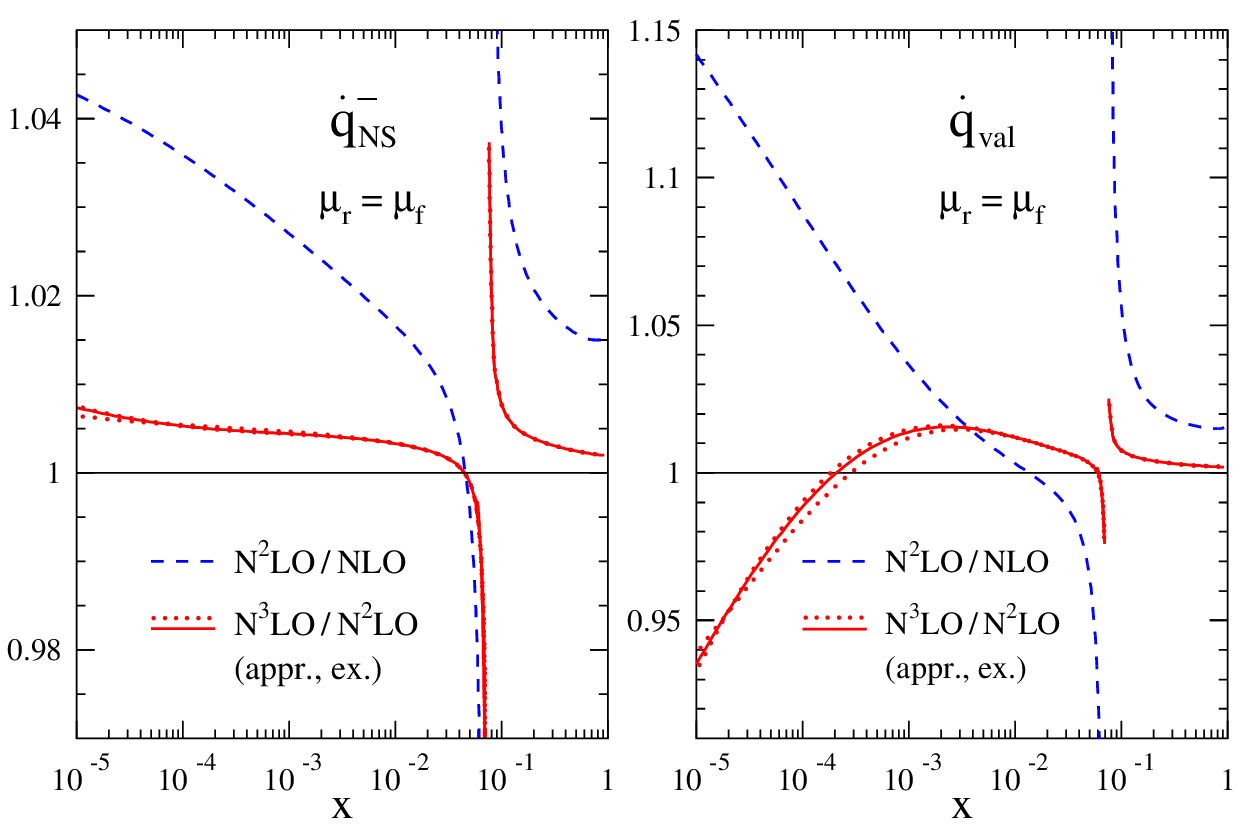,width=15.0cm,angle=0}}
\vspace{-1mm}
\caption{\label{fig:dqns-vx} \small
As the right part of fig.~\ref{fig:pns3p}, but for the expansion of the
splitting functions $P_{\rm ns}^{\,-}$ (left) and $P_{\rm ns}^{\,\rm v}$~%
(right).
}
\end{figure}

\vspace*{0.5mm}
The effects of the three-loop and four-loop splitting functions on the
logarithmic derivatives
$\dot{q}_{\rm ns}^{\, a} \equiv d \ln q_{\rm ns}^{\, a}/ d\ln \mu^{2}$,
$a = \pm,{\rm v}$, are illustrated in the right panel of fig.~\ref{fig:pns3p} 
and in fig.~\ref{fig:dqns-vx} for the default choice $\mu \equiv \mu_{\!f}^{} 
= \mu_r^{}$ of the factorization and renormalization scales already indicated
in eq.~(\ref{eq:evol}).
In all three cases we employ, as already in ref.~\cite{Moch:2004pa},
the same schematic model distribution 
\beq
\label{eq:NSinp}
  xq_{\rm ns}^{\pm,\rm v}(x,\mu_{0}^{\,2}) \; = \;
  x^{\, 0.5} (1-x)^3 
\eeq
and an order-independent value
\beq
\label{eq:asref}
  \als (\mu_0^{\,2}) \; = \; 0.2
\eeq
for the strong coupling constant which corresponds to a scale $\mu_0^{\,2}$
in the range $25\ldots 50$ GeV$^2$. 
This facilitates a direct comparison of effects of the various contributions 
to the splitting functions.


\vspace*{0.5mm}
For the scale dependence of the flavour asymmetries $q_{\rm ns}^{\pm}$,
the relative N$^3$LO corrections are well below 1\% over the full range
shown in the figures, except very close to the sign changes at 
$x \approx 0.07$. This represents a clear accuracy improvement over the
corresponding N$^2$LO results -- for the much larger NLO effects the
reader is referred to figs.~6 and 7 of ref.~\cite{Moch:2004pa}.
The situation is different for the total valence distribution 
$q_{\rm val} \equiv q_{\rm ns}^{\,\rm v}$ at small $x$, due to the
effects of the additional contributions $P_{\!\rm ns}^{(n)\rm s}$ at $n=2$
and $n=3$. These are suppressed at large $x$, but much larger than their
$P_{\!\rm ns}^{\:\!-}$ counterparts at small $x$, see fig.~5 of 
ref.~\cite{Moch:2004pa} and fig.~3 of ref.~\cite{Gehrmann:2026qbl}.

\vspace*{0.5mm}
We have first used the {\sc Fortran} code of ref.~\cite{Gehrmann:2001pz} for 
the numerical evaluation of harmonic polylogarithms (HPLs) $H_{\vec{w}}(x)$
\cite{Remiddi:1999ew}, 
extended to weight-6 \cite{Gehrmann:pc} for use in ref.~\cite{Moch:2017uml},
for the numerical evaluation of the exact results for $P_{\!\rm ns}^{(3)a}(x)$.
Using its numerical output, we have then extended our parametrizations from 
the fermionic and large-$\nc$ contributions 
\cite{Davies:2016jie,Moch:2017uml,Kniehl:2025ttz} to the complete
non-singlet splitting functions \cite{Gehrmann:2026qbl}.
Using the abbreviations
\bea
\label{eq:logs}
  {\cal D}_{\,0} \: = \: 1/(1\!-\!x)_+ \: ,
  \quad x_1 \: = \: 1\!-\!x \: ,
  \quad L_1 \: = \: \ln (1\!-\!x) \: ,
  \quad L_0 \: = \: \ln x 
\; ,
\eea
the N$^3$LO contribution to $P_{\!\rm ns}^{\,+}(x)$ in QCD 
can be approximated by
\bea
\label{eq:Pns3p}
{\lefteqn{ 
P_{\rm ns}^{\,(3)+}(x) 
 \:\:=\:\: 
}}
\nn \\ && \mbox{} \phantom{+} 
%
 \big\{ \, 2.070235 \cdot 10^{\,4}
 - 5.171916 \cdot 10^{\,3}\, \nf
 + 1.955772 \cdot 10^{\,2}\, \nfs
 + 3.272344 \cdot 10^{\,0}\, \nft \big\} \, {\cal D}_{\,0} 
\nn \\ && \mbox{}
 + \big\{ \,
   23393.508
 - 5550.044\,\nf
 + 193.8591\,\nfs
 + 3.014982\,\nft \big\}\,\delta(x_1) 
\nn \\[-1mm] && \mbox{}
 + 25000\, \Big( \, x_1 \big(3.8985 + 2.5079\,x - 0.56226\,x^{2} 
 + 0.41604\,x^{3}\big) - 1.8664\,x_1 L_1 
\nn \\[-2mm] && \mbox{}
 - 1.8636\,L_0 L_1 + 0.17547\,x_1 L_1^{2} + 0.5684\,x L_0 
 + 0.63438\,x L_0^{2} + 0.06534\,x L_0^{3} \Big) 
\nn \\[-1mm] && \mbox{}
 + 5.374286\cdot 10^{\,4}\,L_0 + 1.580664\cdot 10^{\,4}\,L_0^{2} 
 + 2.641883\cdot 10^{\,3}\,L_0^{3} + 2.698936\cdot 10^{\,2}\,L_0^{4} 
\nn \\ && \mbox{}
 + 1.316872\cdot 10^{\,1}\,L_0^{5} + 3.511660\cdot 10^{-1}\,L_0^{6} 
 - 3.1828934\cdot 10^{\,4} + 1.6950937\cdot 10^{\,4}\,L_1
\nn \\ && \mbox{\hspn}
 + \nf \* \Bigl\{
 \, 25000 \,\Big( \, x_1 \big(-0.74096 + 0.59892\,x - 0.15132\,x^{2} 
 + 0.10925\,x^{3}\big) + 1.0988\,x_1 L_1
\nn \\[-2mm] && \mbox{}
 + 1.1103\,L_0 L_1 - 0.01060\,x_1 L_1^{2} + 1.0576\,x L_0 
 + 0.02492\,x L_0^{2} + 0.00670\,x L_0^{3} \Big)
\nn \\[-1mm] && \mbox{}
 - 8.700346\cdot 10^{\,3}\,L_0 - 2.658355\cdot 10^{\,3}\,L_0^{2} 
 - 3.891653\cdot 10^{\,2}\,L_0^{3} 
\nn \\[-1mm] && \mbox{}
 - 3.134156\cdot 10^{\,1}\,L_0^{4} 
 - 1.053498\,L_0^{5} + 8.420314\cdot 10^{\,3} - 2.741830\cdot 10^{\,3}\,L_1
 \Bigr\}
\nn \\ && \mbox{\hspn}
 + \nfs \* \Bigl\{
 \, 250\,\Big( \, x_1 \big(3.0008 + 0.8619\,x - 0.12411\,x^{2} 
 + 0.31595\,x^{3}\big) - 0.37529\,x L_0
\nn \\[-2mm] && \mbox{}
 - 0.21684\,x\,L_0^{2} - 0.02295\,x L_0^{3} + 0.03394\,x_1 L_1 
 + 0.40431\,L_0 L_1 \Big) 
\nn \\[-1mm] && \mbox{}
 + 3.930056\cdot 10^{\,2}\,L_0 + 1.125705\cdot 10^{\,2}\,L_0^{2} 
 + 1.652675\cdot 10^{\,1}\,L_0^{3} 
\nn \\[-1mm] && \mbox{}
 + 7.901235\cdot 10^{-1}\,L_0^{4} - 3.760092\cdot 10^{\,2} 
 + 2.668861\cdot 10^{\,1}\,L_1
 \Bigr\}
\\[-1mm] && \mbox{\hspn}
 + \nft \* \Bigl\{ \,
 - 2.426296 - 0.8460488\,x 
 + \big( 0.5267490\,x_1^{-1} - 3.687243
 + 3.160494\,x\big) \,L_0 
\nn \\[-2mm] && \mbox{}
 - \big( 1.316872\,(x_1^{-1}+ 0.1) - 1.448560\,x \big)\,L_0^{2}
 - \big( 0.2633745\,x_1^{-1} - 0.131687\,(1+x) \big)\,L_0^{3}
 \, \Bigr\}
\nn
\; .
\eea
The $\nft$ part, which is the Mellin inverse of an old leading large-$\nf$ 
result in $N$-space \cite{Gracey:1994nn}, is exact up to the rounding of 
the coefficients.
In all other parts, the coefficients of the form $a \cdot 10^{\,b}$ are 
exact up to a rounding of $a$, while the five-digit coefficients in the 
larger round brackets have been fitted to the exact values of 
$P_{\rm ns}^{\,(3)+}(x)$ in the range $10^{\,-6} \leq x \leq 1-10^{\,-6}$.
Finally the coefficients of $\delta (1\!-\!x)$ in the second line of
eq.~(\ref{eq:Pns3p}) have been slightly adjusted from their exact values 
\cite{Gehrmann:2026qbl}, by $-0.282$, $-0.001$ and $0.0037$ for the $\nfz$, 
$\nfo$ and $\nfs$ contributions, for a maximal accuracy of the second 
moment and convolutions with the quark distributions. 

\vspace*{0.5mm}
A corresponding accurate parametrization for $P_{\!\rm ns}^{\,-}(x)$ 
is given by
\bea
\label{eq:Pns3m}
{\lefteqn{
P_{\rm ns}^{\,(3)-}(x)
 \:\:=\:\:
}}
\nn \\ && \mbox{} \phantom{+}
%
 \big\{ \, 2.070235 \cdot 10^{\,4}
 - 5.171916 \cdot 10^{\,3}\, \nf
 + 1.955772 \cdot 10^{\,2}\, \nfs
 + 3.272344 \cdot 10^{\,0}\, \nft \big\} \, {\cal D}_{\,0}
\nn \\ && \mbox{}
 + \big\{ \,
   23393.102
 - 5550.030\,\nf
 + 193.8559\,\nfs
 + 3.014982\,\nft \big\}\,\delta(x_1)
\nn \\[-1mm] && \mbox{} \,
 +  25000 \,\Big( \, x_1 \big(4.0424 + 2.2915\,x - 0.90975\,x^{2} 
 + 0.84854\,x^{3}\big) - 0.81082\,x_1 L_1 
\nn \\ && \mbox{}
 - 0.88065\,L_0 L_1 + 0.19038\,x_1 L_1^{2} + 0.43116\,x L_0 
 + 0.12820\,x L_0^{2} + 0.00391\,x L_0^{3} \Big)
\nn \\[-1mm] && \mbox{}
 + 5.615655\cdot 10^{\,4}\,L_0 + 1.970952\cdot 10^{\,4}\,L_0^{2} 
 + 3.213435\cdot 10^{\,3}\,L_0^{3} + 2.981618\cdot 10^{\,2}\,L_0^{4} 
\nn \\ && \mbox{}
 + 1.851523\cdot 10^{\,1}\,L_0^{5} + 9.964335\cdot 10^{-1}\,L_0^{6} 
 - 3.1828934\cdot 10^{\,4} + 1.6950937\cdot 10^{\,4}\,L_1
\nn \\[1mm] && \mbox{\hspn}
 + \nf \* \Bigl\{ \,
 25000\,\Big( \, x_1 \big(-0.80350 + 0.72278\,x - 0.16905\,x^{2} 
 + 0.11719\,x^{3}\big) + 1.3324\,x_1 L_1 
 \nn \\[-2mm] && \mbox{}
 + 1.3990\,L_0 L_1 - 0.01710\,x_1 L_1^{2} + 1.2984\,x L_0  
 + 0.00738\,x\,L_0^{2} + 0.00487\,x L_0^{3} \Big)
\nn \\[-1mm] && \mbox{}
 - 1.023783\cdot 10^{\,4}\,L_0 - 3.297010\cdot 10^{\,3}\,L_0^{2} 
 - 4.709831\cdot 10^{\,2}\,L_0^{3} - 3.460082\cdot 10^{\,1}\,L_0^{4} 
\nn \\[-1mm] && \mbox{}
 - 9.744856\cdot 10^{-1}\,L_0^{5} + 8.420314\cdot 10^{\,3} 
 - 2.741830\cdot 10^{\,3}\,L_1
\Bigr\}
\nn \\ && \mbox{\hspn}
 + \nfs \* \Bigl\{ \, 
 250 \,\Big( \, x_1 \big(3.2206 + 1.7507\,x + 0.13281\,x^{2} 
 + 0.45969\,x^{3}\big) + 1.5641\,x L_0
\nn \\[-2mm] && \mbox{}
  - 0.37902\,x\,L_0^{2} - 0.03248\,x L_0^{3} + 2.7511\,x_1 L_1 
  + 3.2709\,L_0 L_1\Big) 
\nn \\[-1mm] && \mbox{}
  + 4.378810\cdot 10^{\,2}\,L_0 + 1.282948\cdot 10^{\,2}\,L_0^{2} 
  + 1.959945\cdot 10^{\,1}\,L_0^{3}
\nn \\[-1mm] && \mbox{}
  + 9.876543\cdot 10^{-1}\,L_0^{4} 
  - 3.760092\cdot 10^{\,2} + 2.668861\cdot 10^{\,1}\,L_1
\Bigr\}
\\ && \mbox{\hspn}
 + \nft \* \Bigl\{ \,
 - 2.426296 - 0.8460488\,x
 + \big( 0.5267490\,x_1^{-1} - 3.687243
 + 3.160494\,x\big) \,L_0
\nn \\[-2mm] && \mbox{}
 - \big( 1.316872\,(x_1^{-1}+ 0.1) - 1.448560\,x \big)\,L_0^{2}
 - \big( 0.2633745\,x_1^{-1} - 0.131687\,(1+x) \big)\,L_0^{3}
 \, \Bigr\}
\nn
\; ,
\eea
where the $\nft$ part is the same as in eq.~(\ref{eq:Pns3p}), and the
$\delta (1\!-\!x)$ adjustments, fixed by the numerical determination of
the first moment, are $-0.688$, $-0.015$ and $0.0095$. 

\vspace*{0.5mm}
Finally the four-loop contribution to $P_{\!\rm ns}^{\,\rm s}(x)$ 
can be approximated by
\bea
\label{eq:Pns3s}
{\lefteqn{ 
P_{\rm ns}^{\,(3)\rm s}(x) 
 \:\:=\:\: 
}}
\nn \\ && \mbox{}
 \nf \* \Bigl\{ \,
 250 \,\Big( \, x_1 \big(34.060 - 0.84735\,x - 6.4973\,x^{2} + 1.1689\,x^{3} 
 - 0.32905\,x L_0 \big)
\nn \\[-2mm] && \mbox{}
 - 4.6473\,x L_0^{2} - 1.7459\,x L_0^{3} 
 - x_1^{2}\,L_1\,\big(1.2384 + 0.13345\,L_1 \big)\Big) 
 + 7.729670\cdot 10^{\,3}\,L_0 
\nn \\[-1mm] && \mbox{}
 + 2.306325\cdot 10^{\,3}\,L_0^{2} - 1.629627\cdot 10^{\,2}\,L_0^{3} 
 - 1.694370\cdot 10^{\,1}\,L_0^{4} + 1.580247\,L_0^{5}
\nn \\[-1mm] && \mbox{}
 - 1.876543\,L_0^{6} - 2.743332\cdot 10^{\,2}\,x_1 L_1 
 + 1.164936\cdot 10^{-1}\,x_1 L_1^{2}
 \Bigr\}
\nn \\ && \mbox{\hspn}
 + \nfs \* \Bigl\{ \,
 250\,\Big( \, x_1 \big(-4.7702 + 1.7922\,x + 0.64510\,x^{2} - 0.15489\,x^{3} 
 - 0.64749\,x\,L_0\big)
\nn \\[-2mm] && \mbox{}
 + 0.83597\,x L_0^{2} + 0.10964\,x L_0^{3} 
 + x_1^{2} L_1\,\big(0.03048 + 0.02146\,L_1\big)\Big) 
 - 6.473971\cdot 10^{\,2}\,L_0 
\nn \\[-1mm] && \mbox{}
 - 6.641219\cdot 10^{\,1}\,L_0^{2} - 5.353347\,L_0^{3} - 5.925926\,L_0^{4} 
 - 3.950617\cdot 10^{-1}\,L_0^{5}
\nn \\[-1mm] && \mbox{}
 + 1.970002\cdot 10^{\,1}\,x_1 L_1 - 3.435474\,x_1 L_1^{2}
\Bigr\}
\; .
\eea
Also here the small-$x$ and large-$x$ coefficients written in the form
$a \!\cdot\! 10^{\,b}$ are exact up to a rounding of $a$ to seven significant 
digits.

\section{The singlet cases}
\setcounter{equation}{0}

\subsection{General properties}

\vspace*{-0.5mm}
So far, all-$N$ results for the N$^3$LO singlet anomalous dimensions have 
been obtained only for the $\nft$ terms, the complete $\nfs$ contributions to 
$\gamma_{\rm ps}^{(3)}$ (pure singlet, see eq.~(\ref{eq:Pdecomp})) and 
$\gamma_{\rm gq}^{(3)}$, and the $\cfs \nfs$ part of $\gamma_{\rm gg}^{(3)}$
\cite{Davies:2016jie,Gehrmann:2023cqm,Falcioni:2023tzp,Kniehl:2026eij}.
In all cases, except for the $\nfz$ parts of $\gamma_{\rm gq}^{(3)}$
and the $\nfo$ parts of $\gamma_{\rm qg}^{(3)}$, the non-rational 
($\zeta$-function) contributions are completely known 
\cite{Davies:2017hyl,Falcioni:2023luc,Falcioni:2023vqq,Falcioni:2024xyt,%
Falcioni:2024qpd,Falcioni:2025hfz,Kniehl:2026eij}.
Already these results show several features that did either not occur 
at three loops and$/$or that transcend the structure (\ref{eq:gamStr}) 
extended, as already for $\gamma_{\rm ns}^{\,\rm s}$, by terms with powers 
of $(N-1)^{-1}$ and~$(N+2)^{-1}$.

\vspace*{0.5mm}
All singlet anomalous dimensions except $\gamma_{\rm ps}^{(3)}$ -- where
terms with $k = 0$ in eq.~(\ref{eq:gamStr}) occur for the first time --
include $\zeta$-function terms with $N$ in the numerator, 
in the form $(N-1)$ and $(N+2)$ for $\gamma_{\rm qg}^{(3)}$ and 
$\gamma_{\rm gq}^{(3)}$, respectively, 
and with $N(N+1)$ for $\gamma_{\rm gg}^{(3)}$.
Where both the $\z5$ and $\z3$ terms are fully known, these numerator
structures are multiplied by the $\zeta$-function part of the function
\bea
  \label{fNfct}
     f(N) &\!=\!&
        5 \* \z5
       + 4 \:\!\* \z3 \:\!\* \S(-2)
       - 2 \* \S(-5)
       - 4 \:\!\* \S(-2,-3)
       + 8 \* \S(-2,-2,1)
       + 4 \:\!\* \S(3,-2)
       - 4 \:\!\* \S(4,1)
       + 2 \* \S(5)
\nn \\ 
         &\!=\!&
         5\:\!\* \z5
       - 2\:\!\* \z3\,\*\BS(2)
       - \BS(2,1,2)
       + \BS(3,1,1)
\; ,
\eea
where, as usual, the sums are evaluated at the argument $N$.
$f(N)$ behaves as $1/N^2$ at large-$N$. 
This function was first encountered, also with positive powers on $N$,
in the three-loop coefficient functions for inclusive DIS
\cite{Vermaseren:2005qc,Moch:2008fj}. 
As expected from the $\z5$ and $\z3$ terms known before, it occurs as 
$S_1^{\,2} f(N)$ as the `wrapping correction' \cite{Bajnok:2008qj} 
in the $\nfz$ parts \cite{Gehrmann:2026qbl} of $\gamma_{\rm ns}^{(n)\pm}(N)$.
There is a second combination of sums that occurs with $N(N+1)$ in 
$\gamma_{\rm gg}^{(3)}$, and without this prefactor in $\gamma_{\rm ps}^{(3)}$ 
\cite{Falcioni:2023luc,Kniehl:2026eij}:
\beq
   \widetilde{f}(N) \:=\: \S(3) - 2 \S(-3) + 4 \S(2,1) 
\quad \mbox{ with } \quad
   \widetilde{f}(N) \ra N^{-2} ( 2 \ln N - 1/2 ) 
\;\; \mbox{ for } \;\;
   N \ra \infty
\; .
\eeq

Another structure that does not occur in the anomalous dimensions up to 
three loops \cite{Moch:2004pa,Vogt:2004mw} (note the argument of the sum) is 
\cite{Gehrmann:2023cqm,Falcioni:2023tzp}
\beq
\label{ftNfct}
   x^{\:\!-2}\, H_{-1,0}(x) \:=\: 
   x^{\:\!-2}\, [ \,\ln(x) \ln(1+x) + \mbox{Li}_2(-x) ]
   \quad \Leftrightarrow \:\:
   (N-2)^{-2} \S(2)(N\!-\!2) 
\; .
\eeq
Its $N=2$ moment is $-3/2\,\z3$, hence it leads to a $\delta_{N,2}$ term
in the analytical $N$-space expressions with $\z3$ occurring in all four-loop
singlet anomalous dimensions, but not in the non-singlet cases. 
The function (\ref{ftNfct}) does occur already in two-loop coefficient
functions in DIS, first calculated in 
refs.~\cite{vanNeerven:1991nn,Zijlstra:1991qc} and,
incidentally, presented at what became the first in this series of meetings.

\subsection{Numerical results}

\begin{figure}[p]
\vspace{-4mm}
\centerline{\hspace*{-2mm}\epsfig{file=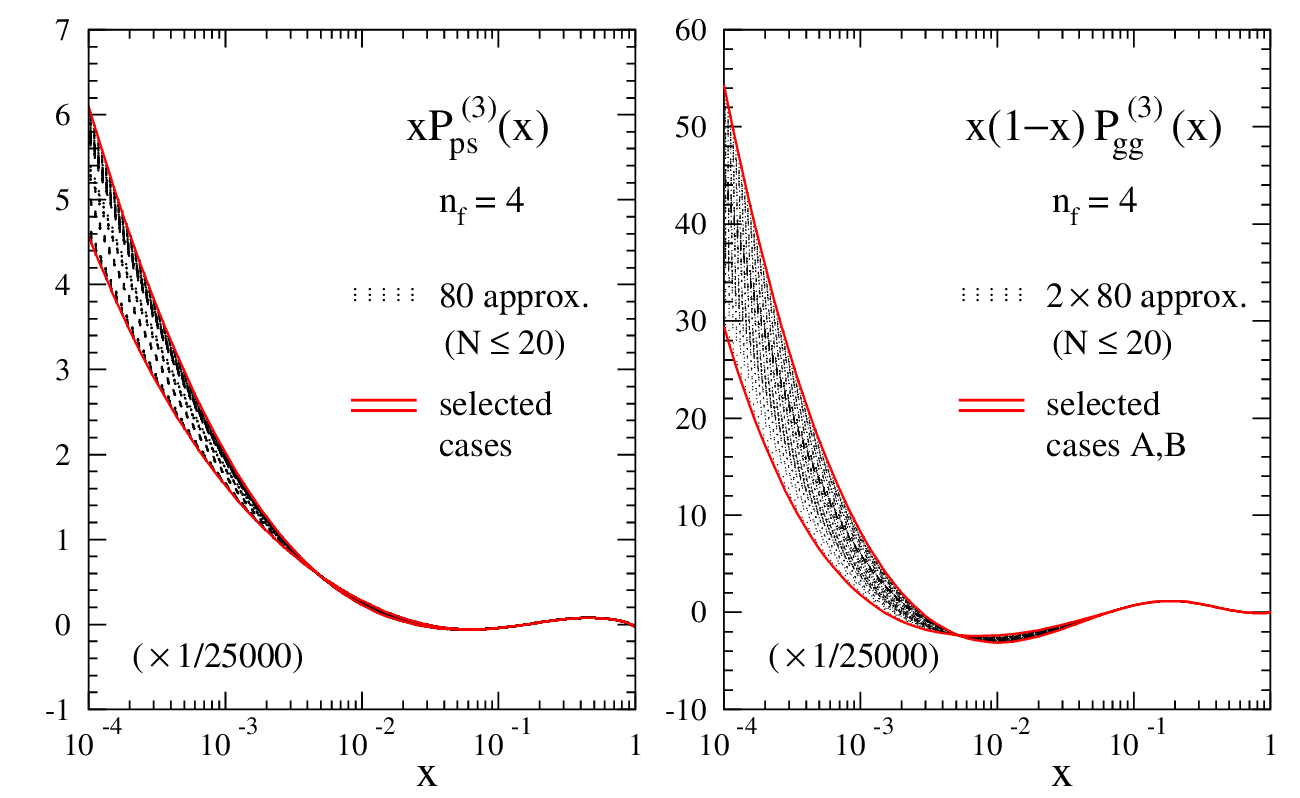,width=15.5cm,angle=0}}
\vspace{-2mm}
\caption{\label{fig:pii3ab4} \small
Approximations for the 4-loop contributions to the pure-singlet (ps) and
gluon-gluon splitting functions in eq.~(\ref{eq:Pdecomp}) for $\nf = 4$ light
flavours, as constructed in refs.~\cite{Falcioni:2023luc,Falcioni:2024qpd}.
The curves have been divided by $25000 \simeq (4\pi)^4$, they thus correspond
to an expansion in powers of $\als$ instead of $\als / (4\pi)$ as in 
eq.~(\ref{eq:Pexp}).
}
\end{figure}

\begin{figure}[p]
\vspace{-4mm}
\centerline{\hspace*{-3mm}\epsfig{file=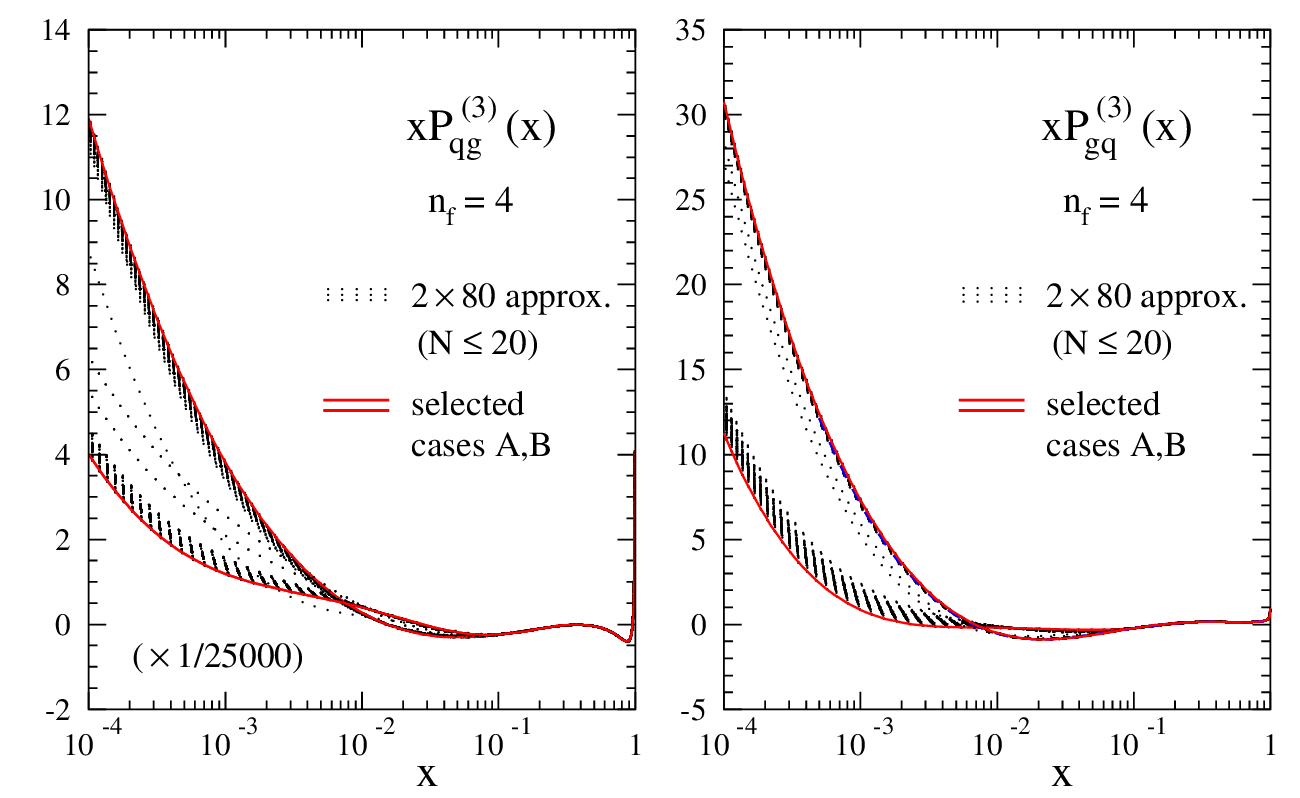,width=15.5cm,angle=0}}
\vspace{-2mm}
\caption{\label{fig:pij3ab4} \small
As fig.~\ref{fig:pii3ab4}, but for the gluon-to-quark (qg) and 
quark-to-gluon (gq) splitting functions as approximated in 
refs.~\cite{Falcioni:2023vqq,Falcioni:2025hfz}. 
}
\end{figure}

\vspace*{-0.5mm}
In refs.~\cite{Falcioni:2023luc,Falcioni:2023vqq,Falcioni:2024xyt,%
Falcioni:2024qpd} we have computed the even-$N$ moments of all four four-loop
singlet splitting functions (\ref{eq:Pdecomp}) to $N \!=\! 20$ via operator
matrix elements using {\sc Forcer} \cite{Ruijl:2017cxj}, and used these
moments, supplemented by small-$x$ and large-$x$ endpoint information
\cite{Henn:2019swt,vonManteuffel:2020vjv,Davies:2022ofz,%
Catani:1989sg,Catani:1994sq,Fadin:1998py,Ciafaloni:1998gs,Ciafaloni:2005cg,%
Ciafaloni:2006yk,Soar:2009yh,Almasy:2010wn}, 
to construct approximate $x$-space expressions that are sufficiently
accurate over a rather wide range in~$x$, but unavoidably suffer from large 
uncertainties at $x \lsim 10^{-3}$. 
They have been validated, and for $P_{\!\rm qg}^{\,(3)}(x)$ slightly shifted,
by comparing the resulting error band at $N=22$ with the exact values in 
ref.~\cite{Falcioni:2025hfz}. The resulting uncertainty bands are shown in 
figs.~\ref{fig:pii3ab4} and \ref{fig:pij3ab4} for four light flavours.


Due to the interplay of large-$N$ constraints sampling high values 
of $x$ in eq.~(\ref{eq:PvsGam}), the smooth intermediate-$x$ functions,
and the small-$x$ predictions, the latter depend on how well the 
large-$x$ behaviour of the splitting functions is known beforehand. 
This situation is most favourable for pure-singlet case, with unknown 
coefficients of $(1-x) \ln^{\,k\!} x$, $k=1,\,2$, 
somewhat worse for the gluon-gluon case 
-- before refs.~\cite{Gehrmann:2026qbl,Moch:2026qsw}, due to a very small 
but not irrelevant residual uncertainty of the $\delta (1-x)$ contribution -- 
and considerably worse for the off-diagonal cases, where the $(1-x)^0 
\ln^{\,k'\!} (1-x)$ are covered by the present generalized large-$x$ 
resummation only for $k'>3$ \cite{Soar:2009yh,Almasy:2010wn}.

\begin{figure}[tbh]
\vspace{1mm}
\centerline{\hspace*{1mm}\epsfig{file=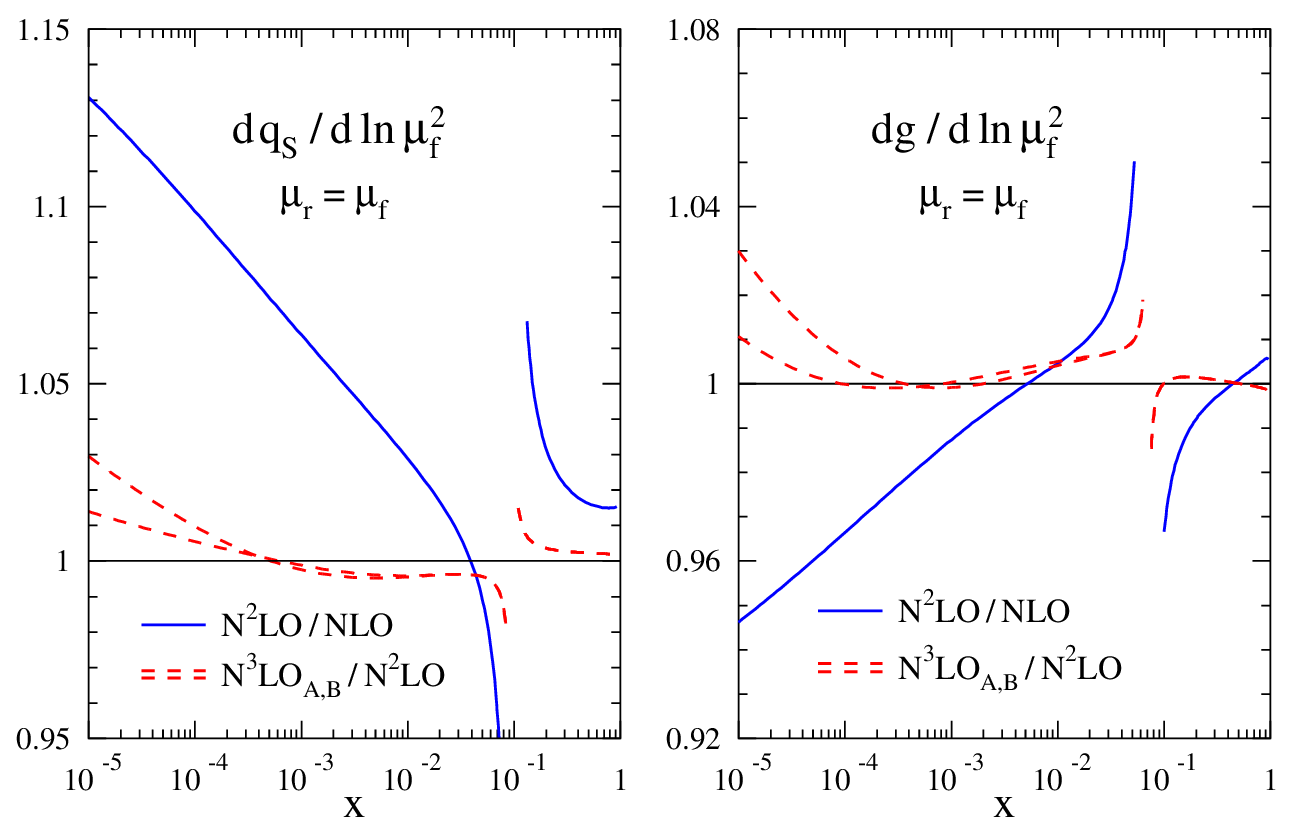,width=15.5cm,angle=0}}
\vspace{-2mm}
\caption{\label{fig:dsgn3lo} \small
The relative N$^2$LO and N$^3$LO corrections to the scale derivatives of
singlet quark (left panel) and gluon (right panel) PDFs at a standard 
reference scale $\mu_0^{}$ with $\nf=4$ and $\als(\mu_0^{\,2}) = 0.2$, 
as obtained using the selected approximations in figs.~\ref{fig:pii3ab4} 
and \ref{fig:pij3ab4}.
Note that that $dq_{\rm s}^{} / d\ln \mu^2$ and $dg / d\ln \mu^2$ change 
sign close to $x=0.1$ and $x=0.07$, respectively, which leads to 
irrelevant singularities in the relative corrections.
}
\end{figure}

\vspace*{0.5mm}
In order to illustrate the size and uncertainties of the present N$^3$LO 
contributions to the evolution of the singlet PDFs, we choose, as for the 
previous order in ref.~\cite{Vogt:2004mw}, $\als (\mu_{0}^{\,2}) =\, 0.2$, 
$\nf = 4$, and the sufficiently realistic input distributions
\bea
\label{eq:SGinp}
  xq_{\rm s}^{}(x,\mu_{0}^{\,2}) &\! = \!&
  0.6\: x^{\, -0.3} (1-x)^{3.5}\, \big( 1 + 5.0\: x^{\, 0.8\,} \big)
\: ,
\nn \\
  xg(x,\mu_{0}^{\,2}) &\! = \!&
  1.6\: x^{\, -0.3} (1-x)^{4.5}\, \big( 1 - 0.6\: x^{\, 0.3\,} \big)
\; .
\label{eq:qg-shape}
\eea
The resulting relative N$^2$LO and N$^3$LO effects on the scale derivative
at this reference point are shown in fig.~\ref{fig:dsgn3lo}. Due to the
Mellin convolutions with $q_{\rm s}$ and $g$ in eq.~(\ref{eq:SGinp}), 
the uncertainties of the 4-loop splitting functions become relevant only
at $x$-values below about $10^{-4}$. 
The N$^3$LO corrections are generally small, reaching $(2 \pm 1)\%$ at 
$x = 10^{-5}$. The higher-order effect will be smaller at higher scales and
larger at smaller scales.

\section{Concluding remarks}

\vspace*{-1mm}
The recently completed N$^3$LO non-singlet splitting functions 
$P_{\!\rm ns}^{(3)}$ \cite{Gehrmann:2026qbl} agree with all 
previous numerical and analytical expectations, see also
ref.~\cite{Moch:2026qsw}, 
except for next-to-next-to-leading small-$x$ logarithms, 
where the $\nfz$ terms with $\z2$ differ, beyond
the limit of a large number of colours, from the predictions in
ref.~\cite{Davies:2022ofz}.
The new four-loop structure indicated by these differences and
the single-log terms in the $\nfz$ quartic-Casimir contributions
may also affect related predictions for the coefficient functions
in inclusive DIS \cite{Davies:2022ofz} and related quantities in
semi-inclusive $e^+ e^-$ annihilation~\cite{Vogt:2011jv}.

\vspace*{0.5mm}
The N$^3$LO singlet splitting functions are only known in an 
approximate form \cite{Falcioni:2025hfz} which, however, should
be sufficiently accurate for most practical applications.
We expect that also these functions will be computed rather soon
along the lines of ref.~\cite{Gehrmann:2026qbl}, hence large
efforts for incremental improvements upon 
ref.~\cite{Falcioni:2025hfz} are not called for.
In the absence of complete results, though, the determination
of all $x^{-1} \ln x$ terms would be a game changer. 
In this context it seems remarkable that the 1994 paper 
\cite{Catani:1994sq} still appears to represent the state 
of the art for $P_{\!\rm ps,\,qg}^{(3)}$ at small $x$.

\vspace*{2mm}
A {\sc Form} file of the expansion coefficients in 
eqs.~(\ref{eq:Pp1smallx}) -- (\ref{eq:Ps3smallx})
and {\sc Fortran} files of the parametrizations 
(\ref{eq:Pns3p}) -- (\ref{eq:Pns3s}), as well their exact counterparts,
can be obtained from the preprint server http://arXiv.org,
in a ready-to-use form analogous to the codes distributed
with refs.~\cite{Moch:2004pa,Vogt:2004mw,Moch:2017uml}.
They are also available, for now, from the authors upon request.

\subsection*{Acknowledgements}

\vspace*{-1mm}
This work has been supported by
the Deutsche Forschungsgemeinschaft (DFG), 
Research Unit FOR 2926, project number 409651613;
the European Research Council (ERC) Advanced Grant 101095857;
the UK Science and Technology Facilities Council (STFC), 
Consolidated Grant ST/X000699/1;
and by a Research Award of the Alexander-von-Humboldt Foundation.


\vspace*{-1mm}
\begin{thebibliography}{99}
\addtolength{\baselineskip}{-0.4mm}
\setlength{\itemsep}{1mm}
\vspace*{-1mm}

\bibitem{Moch:2004pa}
S.~Moch, J.A.M.~Vermaseren and A.~Vogt,
  \emph{{The three loop splitting functions in QCD:\\ the non-singlet case}},
  \href{https://dx.doi.org/10.1016/j.nuclphysb.2004.03.030}
  {Nucl.\ Phys.\ B688 (2004) 101}
  [\href{https://arxiv.org/abs/hep-ph/0403192}{{\tt hep-ph/0403192}}]

\bibitem{Vogt:2004mw}
A.~Vogt, S.~Moch and J.A.M. Vermaseren,
  \emph{{The three-loop splitting functions in QCD:\\ the singlet case}}
  \href{https://doi.org/10.1016/j.nuclphysb.2004.04.024}
  {Nucl.\ Phys.\ B691 (2004) 129},
  [\href{https://arxiv.org/abs/hep-ph/0404111}{{\tt hep-ph/0404111}}]

\bibitem{Vermaseren:2005qc}
J.A.M. Vermaseren, A.~Vogt and S.~Moch,
  \emph{{The third-order QCD corrections to deep-inelastic scattering by
  photon exchange}}, \href{https://doi.org/10.1016/j.nuclphysb.2005.06.020}
 {Nucl.\ Phys.\ B724 (2005) 3}
 [\href{https://arxiv.org/abs/hep-ph/0504242}{{\tt hep-ph/0504242}}]

\bibitem{Moch:2008fj}
S.~Moch, J.A.M. Vermaseren and A.~Vogt,
  \emph{{Third-order QCD corrections to the charged- current structure function
  $F_3$}}, \href{https://doi.org/10.1016/j.nuclphysb.2009.01.001}
 {Nucl.\ Phys.\ B813 (2009) 220}
 [\href{https://arxiv.org/abs/0812.4168}{{\tt arXiv:0812.4168}}]

\bibitem{Currie:2018fgr}
J.~Currie et al,
  \emph{{N$^{3}$LO corrections to jet production in deep inelastic scattering 
  using the Projection-to-Born method}}, 
  \href{https://doi.org/10.1007/JHEP05(2018)209}{JHEP 05 (2018) 209}
 [\href{https://arxiv.org/abs/1803.09973}{{\tt arXiv:1803.09973}}]

\bibitem{Gehrmann:2018odt}
T.~Gehrmann, A.~Huss, J.~Niehues, A.~Vogt and D.~M.~Walker,
  \emph{{Jet production in charged-current deep-inelastic scattering to 
  third order in QCD}},
  \href{https://dx.doi.org/10.1016/j.physletb.2019.03.003}
  {Phys.\ Lett.\ B792 (2019) 182}  \\
  {[\href{https://arxiv.org/abs/1812.06104}{{\tt arXiv:1803.06014}}]}

\bibitem{Anastasiou:2015vya}
C.~Anastasiou, C.~Duhr, F.~Dulat, F.~Herzog, B.~Mistlberger,
 \emph{{Higgs Boson Gluon-Fusion Production in QCD at Three Loops}},
 \href{https://doi.org/10.1103/PhysRevLett.114.212001}{Phys.\ Rev.\ Lett.\
 114 (2015) 212001}
 [\href{https://arxiv.org/abs/1503.06056}{{\tt arXiv:1503.06056}}]

\bibitem{Mistlberger:2018etf}
B.~Mistlberger,
 \emph{{Higgs boson production at hadron colliders at N$^{3}$LO in QCD}},\\ 
 \href{https://doi.org/10.1007/JHEP05(2018)028}{JHEP 05 (2018) 028}
 [\href{https://arxiv.org/abs/1802.00833}{{\tt arXiv:1802.00833}}]

\bibitem{Chen:2021isd}
X.~Chen et al,
 \emph{{Fully Differential Higgs Boson Production to Third Order in QCD}}, \\
 \href{https://doi.org/10.1103/PhysRevLett.127.072002}
 {Phys.\ Rev.\ Lett.\ 127 (2021) 072002}
 [\href{https://arxiv.org/abs/2102.07607}{{\tt arXiv:2102.07607}}]

\bibitem{Ruijl:2016pkm}
B.~Ruijl, T.~Ueda, J.A.M.~Vermaseren, J.~Davies and A.~Vogt,
  \emph{{First Forcer results on deep-inelastic scattering and related 
  quantities}},
  \href{https://doi.org/10.22323/1.260.0071}{PoS LL2016 (2016) 071}
  [\href{https://arxiv.org/abs/1605.08408}{{\tt arXiv:1605.08408}}]

\bibitem{Davies:2016jie}
J.~Davies, A.~Vogt, B.~Ruijl, T.~Ueda and J.A.M. Vermaseren,
 \emph{Large-$\nf$ contributions to the four-loop splitting functions in QCD},
 \href{https://doi.org/10.1016/j.nuclphysb.2016.12.012}{Nucl.\ Phys.\ B915
 (2017) 335}
 [\href{https://arxiv.org/abs/1610.07477}{{\tt arXiv:1610.07477}}]

\bibitem{Moch:2017uml}
S.~Moch, B.~Ruijl, T.~Ueda, J.A.M. Vermaseren, A.~Vogt,
  \emph{{Four-Loop Non-Singlet Splitting Functions in the Planar Limit
  and Beyond}},
  \href{https://doi.org/10.1007/JHEP10(2017)041}{JHEP 10 (2017) 041}
  [\href{https://arxiv.org/abs/1707.08315}{{\tt arXiv:1707.08315}}]

\bibitem{Davies:2017hyl}
J.~Davies and A.~Vogt,
 \emph{{Absence of $\pi^2$ terms in physical anomalous dimensions in DIS: 
 Verification and resulting predictions}},
 \href{https://doi.org/10.1016/j.physletb.2017.11.036}
 {Phys.\ Lett.\ B776 (2018) 189}
 [\href{https://arxiv.org/abs/1711.05267}{{\tt arXiv:1711.05267}}]

\bibitem{Moch:2018wjh}
S.~Moch, B.~Ruijl, T.~Ueda, J.A.M.~Vermaseren and A.~Vogt,
 \emph{{On quartic colour factors in splitting functions and the gluon 
 cusp anomalous dimension}},
 \href{https://dx.doi.org/10.1016/j.physletb.2018.06.017}
 {Phys.\ Lett.\ B782 (2018) 627} 
 [\href{https://arxiv.org/abs/1805.09638}{{\tt arXiv:1805.09638}}]

\bibitem{Vogt:2018miu}
A.~Vogt, F.~Herzog, S.~Moch, B.~Ruijl, T.~Ueda and J.A.M.~Vermaseren,
  \emph{{Anomalous dimensions and splitting functions beyond the 
  next-to-next-to-leading order}},
  \href{https://doi.org/10.22323/1.303.0050}{PoS LL2018 (2018) 050}
  [\href{https://arxiv.org/abs/1808.08981}{{\tt arXiv:1808.08981}}]

\bibitem{Moch:2021qrk}
S.~Moch, B.~Ruijl, T.~Ueda, J.A.M. Vermaseren and A.~Vogt,
  \emph{{Low moments of the four-loop splitting functions in QCD}},
  \href{https://doi.org/10.1016/j.physletb.2021.136853}
  {Phys.\ Lett.\ B825 (2022) 136853}
  [\href{https://arxiv.org/abs/2111.15561}{{\tt arXiv:2111.15561}}]

\bibitem{Falcioni:2022fdm}
G.~Falcioni and F.~Herzog,
  \emph{{Renormalization of gluonic leading-twist operators in covariant 
  gauges}},
  \href{https://doi.org/10.1007/JHEP05(2022)177}{JHEP 05 (2022) 177}
  [\href{https://arxiv.org/abs/2203.11181}{{\tt arXiv:2203.11181}}]

\bibitem{Falcioni:2023luc}
G.~Falcioni, F.~Herzog, S.~Moch and A.~Vogt,
 \emph{{Four-loop splitting functions in QCD \\ - The quark-quark case }},
 \href{https://doi.org/10.1016/j.physletb.2023.137944}
 {Phys.\ Lett.\ B842 (2023) 137944}
 [\href{https://arxiv.org/abs/2302.07593}{{\tt arXiv:2302.07593}}]

\bibitem{Falcioni:2023vqq}
G.~Falcioni, F.~Herzog, S.~Moch and A.~Vogt,
 \emph{{Four-loop splitting functions in QCD\\ -- The gluon-to-quark case}},
 \href{https://doi.org/10.1016/j.physletb.2023.138215}
 {Phys.\ Lett.\ B846 (2023) 138215}
 [\href{https://arxiv.org/abs/2307.04158}{{\tt arXiv:2307.04158}}]

\bibitem{Gehrmann:2023cqm}
T.~Gehrmann, A.~von~Manteuffel, V.~Sotnikov and T.Z.~Yang,
 \emph{{Complete $\nfs$ contributions to four-loop pure-singlet 
 splitting functions}},
  \href{http://dx.doi.org/10.1007/JHEP01(2024)029}{JHEP 01 (2024) 029}
  [\href{https://arxiv.org/abs/2308.07958}{{\tt arXiv:2308.07958}}]

\bibitem{Gehrmann:2023iah}
T.~Gehrmann, A.~von~Manteuffel, V.~Sotnikov and T.Z.~Yang, 
 \emph{{The $\nf \cft$ contribution to the non-singlet splitting function at 
 four-loop order}}, 
  \href{https://dx.doi.org/10.1016/j.physletb.2023.138427}
  {Phys.\ Lett.\ B849 (2024) 138427}
  [\href{https://arxiv.org/abs/2310.12240}{{\tt 2310.12240}}]

\bibitem{Falcioni:2023tzp}
G.~Falcioni, F.~Herzog, S.~Moch, J.A.M~Vermaseren and A.~Vogt,
 \emph{{The double fermionic contribution to the four-loop quark-to-gluon 
 splitting function}},
 \href{https://dx.doi.org/10.1016/j.physletb.2023.138351}
 {Phys.\ Lett.\ B848 (2024) 138351} 
 [\href{https://arxiv.org/abs/2310.01245}{{\tt arXiv:2310.01245}}]

\bibitem{Moch:2023tdj}
S.~Moch, B.~Ruijl, T.~Ueda, J.~Vermaseren, A.~Vogt,
 \emph{{Additional moments and x-space approximations of four-loop splitting 
 functions in QCD}},
 \href{https://dx.doi.org/10.1016/j.physletb.2024.138468}
 {Phys.\ Lett.\ B849 (2024) 138468} 
 [\href{https://arxiv.org/abs/2310.05744}{{\tt arXiv:2310.05744}}]

\bibitem{Falcioni:2024xyt}
G.~Falcioni, F.~Herzog, S.~Moch, A.~Pelloni, and A.~Vogt.
  \emph{Four-loop splitting functions in QCD -- The quark-to-gluon case},
  \href{https://doi.org/10.1016/j.physletb.2024.138906}
  {Phys.\ Lett.\ B856 (2024) 138906}
  [\href{https://arxiv.org/abs/2404.09701}{{\tt arXiv:2404.09701}}]

\bibitem{Falcioni:2024xav}
G.~Falcioni, F.~Herzog, S.~Moch and S.~Van Thurenhout,
  \emph{{Constraints for twist-two alien operators in QCD}},
  \href{https://dx.doi.org/10.1007/JHEP11(2024)080}{JHEP 11 (2024) 080}
  [\href{https://arxiv.org/abs/2409.02870}{{\tt arXiv:2409.02870}}]

\bibitem{Gehrmann:2024ggw}
T.~Gehrmann, A.~von Manteuffel and T.Z.~Yang,
  \emph{{Leading Twist-Two Gauge-Variant Counterterms}},
  \href{https://dx.doi.org/10.22323/1.467.0087}{PoS LL2024 (2024) 087}
  [\href{https://arxiv.org/abs/2409.10303}{{\tt arXiv:2409.10303}}]

\bibitem{Falcioni:2024qpd}
G.~Falcioni, F.~Herzog, S.~Moch, A.~Pelloni, and A.~Vogt.
  \emph{Four-loop splitting functions in QCD -- The gluon-gluon case},
  \href{http://dx.doi.org/10.1016/j.physletb.2024.139194}
  {Phys.\ Lett.\ B860 (2025) 139194}
  [\href{https://arxiv.org/abs/2410.08089}{{\tt arXiv:2410.08089}}]

\bibitem{Kniehl:2025jfs}
B.A.~Kniehl and V.N.~Velizhanin,
  \emph{{Four-loop anomalous dimension of flavor non-singlet twist-two 
  operator of general Lorentz spin in QCD: $\zeta$(3) term}},
  \href{http://dx.doi.org/10.1103/PhysRevLett.134.131901}
  {Phys.\ Rev.\ Lett.\ 134 (2025) 131901}
  [\href{https://arxiv.org/abs/2503.20422}{{\tt arXiv:2503.20422}}]

\bibitem{Kniehl:2025ttz}
B.A.~Kniehl, S.O.~Moch, V.N.~Velizhanin and A.~Vogt,
  \emph{{Flavor Non-singlet Splitting Functions at Four Loops in QCD:
  Fermionic Contributions}},
  \href{https://dx.doi.org/10.1103/hkg5-88hr}{Phys.\ Rev.\ Lett.\ 135
  (2025) 071902}
  [\href{https://arxiv.org/abs/2505.09381}{{\tt arXiv:2505.09381}}]

\bibitem{Falcioni:2025hfz}
G.~Falcioni, F.~Herzog, S.~Moch, A.~Pelloni and A.~Vogt,
  \emph{Additional results on the four-loop flavour-singlet splitting
  functions in QCD},
  \href{https://doi.org/10.1016/j.physletb.2026.140278}
  {Phys.\ Lett.\ B875 (2026) 140278} 
  [\href{https://arxiv.org/abs/2512.10783}{{\tt 2512.10783}}]

\bibitem{Kniehl:2026eij}
B.A.~Kniehl, S.O.~Moch, V.N.~Velizhanin and A.~Vogt,
  \emph{{Four-Loop Gluon Anomalous Dimension of General Lorentz Spin: 
  Transcendental Part}},
  \href{https://doi.org/10.1103/g4l4-t4cv}
  {Phys.\ Rev.\ Lett.\ 137 (2026) 081901}
  [\href{https://arxiv.org/abs/2604.25833}{{\tt arXiv:2604.25833}}]

\bibitem{Gehrmann:2026qbl}
T.~Gehrmann, A.~von Manteuffel, V.~Sotnikov and T.~Yang,
  \emph{{The four-loop non-singlet splitting functions in QCD}},
  \href{https://arxiv.org/abs/2604.09534}{{\tt arXiv:2604.09534}}

\bibitem{Moch:2026qsw}
S.~Moch and A.~Vogt,
  \emph{{Properties and implications of the four-loop non-singlet 
  splitting functions in QCD}},
  \href{https://arxiv.org/abs/2605.03889}{{\tt arXiv:2605.03889}}

\bibitem{Vermaseren:1998uu}
J.A.M. Vermaseren,
  \emph{{Harmonic sums, Mellin transforms and integrals}}, \\
 \href{https://doi.org/10.1142/S0217751X99001032}
 {Int.\ J. Mod.\ Phys.\ A14 (1999) 2037}
 [\href{https://arxiv.org/abs/hep-ph/9806280}{{\tt hep-ph/9806280}}]

\bibitem{Dokshitzer:2005bf}
   Yu.L.~Dokshitzer, G.~Marchesini and G.P.~Salam,
  \emph{{Revisiting parton evolution and the large-x limit}},
  \href{https://dx.doi.org/10.1016/j.physletb.2006.02.023}
  {Phys.\ Lett.\ B634 (2006) 504}
  [\href{https://arxiv.org/abs/hep-ph/0511302}{{\tt hep-ph/0511302}}]

\bibitem{Dokshitzer:2006nm}
  Yu.L.~Dokshitzer and G.~Marchesini,
  \emph{{$N=4$ SUSY Yang-Mills: three loops made simple(r)}},
  \href{https://dx.doi.org/10.1016/j.physletb.2007.01.016}
  {Phys.\ Lett.\ B646 (2007) 189}
  [\href{https://arxiv.org/abs/hep-th/0612248}{{\tt hep-th/0612248}}]

\bibitem{Basso:2006nk}
B.~Basso and G.P.~Korchemsky,
  \emph{Anomalous dimensions of high-spin operators beyond the leading order},
  \href{https://dx.doi.org/10.1016/j.nuclphysb.2007.03.044}
  {Nucl.~Phys.\ B775 (2007) 1}
  [\href{https://arxiv.org/abs/hep-th/0612247}{{\tt hep-th/0612247}}]

\bibitem{Strohmaier:2018tjo}
M.~Strohmaier,
  \emph{Conformal symmetry breaking and evolution equations in 
  Quantum Chromodynamics}, PhD thesis, Regensburg 2018,
  \href{https://epub.uni-regensburg.de/37432/}
  {{\tt doi:10.5283/epub.37432}}

\bibitem{Mitov:2006ic}
A.~Mitov, S.~Moch and A.~Vogt,
  \emph{Next-to-Next-to-Leading Order Evolution of Non-Singlet \\
  Fragmentation Functions},
  \href{https://dx.doi.org/10.1016/j.physletb.2006.05.005}
  {Phys.\ Lett.\ B638 (2006) 61}
  [\href{https://arxiv.org/abs/hep-ph/0604053}{{\tt hep-ph/0604053}}]

\bibitem{Vermaseren:2000nd}
J.A.M. Vermaseren,
  \emph{{New features of FORM}},
 \href{https://arxiv.org/abs/math-ph/0010025}{{\tt math-ph/0010025}}

\bibitem{Kuipers:2012rf}
J.~Kuipers, T.~Ueda, J. Vermaseren and J.~Vollinga,
  \emph{{FORM version 4.0}}, \\
 \href{https://doi.org/10.1016/j.cpc.2012.12.028}
 {Comput.\ Phys.\ Comm.\ 184 (2013) 1453}
 [\href{https://arxiv.org/abs/1203.6543}{{\tt arXiv:1203.6543}}]

\bibitem{Ruijl:2017dtg}
B.~Ruijl, T.~Ueda and J.~Vermaseren,
  \emph{{FORM version 4.2}},
 \href{https://arxiv.org/abs/1707.06453}{{\tt arXiv:1707.06453}}

\bibitem{Vogt:2012gb}
A. Vogt et al.,
  \emph{Progress on double-logarithmic large-x and small-x resummations for 
  (semi-) inclusive hard processes},
 \href{https://dx.doi.org/10.22323/1.151.0004}{PoS LL2012 (2012) 004}
 [\href{https://arxiv.org/abs/1212.2932}{{\tt arXiv:1212.2932}}]

\bibitem{Davies:2022ofz}
J.~Davies, C.H.~Kom, S.~Moch and A.~Vogt,
  \emph{{Resummation of small-x double logarithms in QCD: 
  inclusive deep-inelastic scattering}},
  \href{https://dx.doi.org/10.1007/JHEP08(2022)135}{JHEP 08 (2022) 135}
  [\href{https://arxiv.org/abs/2202.10362}{{\tt arXiv:2202.10362}}]

\bibitem{Ruijl:2017cxj}
B.~Ruijl, T.~Ueda and J.A.M. Vermaseren,
  \emph{{Forcer, a FORM program for the parametric reduction of four-loop 
  massless propagator diagrams}},
 \href{https://doi.org/10.1016/j.cpc.2020.107198}
 {Comput.\ Phys.\ Comm.\ 253 (2020) 107198}
 [\href{https://arxiv.org/abs/1704.06650}{{\tt arXiv:1704.06650}}]

\bibitem{Henn:2019swt}
J.M.~Henn, G.P.~Korchemsky and B.~Mistlberger,
  \emph{{The full four-loop cusp anomalous dimension in $\mathcal{N}=4$ 
  super Yang-Mills and QCD}},
  \href{https://dx.doi.org/10.1007/JHEP04(2020)018}{JHEP 04 (2020) 018}
  [\href{https://arxiv.org/abs/1911.10174}{{\tt arXiv:1911.10174}}]

\bibitem{vonManteuffel:2020vjv}
A.~von~Manteuffel, E.~Panzer and R.M.~Schabinger,
  \emph{{Cusp and collinear anomalous dimensions in four-loop QCD from 
  form factors}},
  \href{https://dx.doi.org/10.1103/PhysRevLett.124.162001}
  {Phys.\ Rev.\ Lett.\ 124 (2020) 162001} 
  [\href{https://arxiv.org/abs/2002.04617}{{\tt 2002.04617}}]

\bibitem{Gehrmann:2001pz}
T.~Gehrmann and E.~Remiddi, 
  \emph{{Numerical evaluation of harmonic polylogarithms}}, \\
  \href{https://dx.doi.org/10.1016/S0010-4655(01)00411-8}
  {Comput.\ Phys.\ Commun.\ 141 (2001) 296}
  [\href{https://arxiv.org/abs/hep-ph/0107173}{{\tt hep-ph/0107173}}]

\bibitem{Remiddi:1999ew}
E.~Remiddi and J.~Vermaseren, 
  \emph{{Harmonic polylogarithms}},
  \href{https://dx.doi.org/10.1142/S0217751X00000367}
  {Int.\ J. Mod.\ Phys.\ A15 (2000) 725} \\
  {[\href{https://arxiv.org/abs/hep-ph/9905237}{{\tt hep-ph/9905237}}]}

\bibitem{Gehrmann:pc}
T. Gehrmann, 
  \emph{{private communication}}

\bibitem{Gracey:1994nn}
J.A.~Gracey,
  \emph{{Anomalous dimension of non-singlet Wilson operators at
  O$(1/\nf)$ in deep inelastic scattering}},
  \href{https://dx.doi.org/10.1016/0370-2693(94)90502-9}
  {Phys.\ Lett.\ B322 (1994) 141}
  [\href{https://arxiv.org/abs/hep-ph/9401214}{{\tt hep-ph/9401214}}]

\bibitem{Becher:2024kmk}
T.~Becher, P.~Hager, S.~Jaskiewicz, M.~Neubert and D.~Schwienbacher,
  \emph{{Factorization Restoration through Glauber Gluons}},
  \href{https://dx.doi.org/10.1103/PhysRevLett.134.061901}
  {Phys.\ Rev.\ Lett.\ 134 (2025) 061901}
  [\href{https://arxiv.org/abs/2408.10308}{{\tt arXiv:2408.10308}}]

\bibitem{Becher:2025igg}
T.~Becher, P.~Hager, S.~Jaskiewicz, M.~Neubert and D.~Schwienbacher,
  \emph{Low-energy theory of jet processes and PDF factorization},
  \href{https://dx.doi.org/10.1007/JHEP01(2026)024}{JHEP 01 (2026) 024}
  [\href{https://arxiv.org/abs/2509.07082}{{\tt arXiv:2509.07082}}]
    
\bibitem{Blumlein:1995jp}
J.~Bl{\"u}mlein and A.~Vogt,
  \emph{{On the behavior of non-singlet structure functions at small $x$}},
  \href{https://dx.doi.org/10.1016/0370-2693(95)01568-X}
  {Phys.\ Lett.\ B370 (1996) 149}
  [\href{https://arxiv.org/abs/hep-ph/9510410}{{\tt hep-ph/9510410}}]

\bibitem{Blumlein:1996aw}
J.~Bl{\"u}mlein, S.~Riemersma and A.~Vogt,
  \emph{The Evolution of unpolarized and polarized structure functions 
  at small $x$},
  \href{https://dx.doi.org/10.1016/S0920-5632(96)90005-5}
  {Nucl.\ Phys.\ B Proc.\ Suppl.\ 51 (1996) 30}
  [\href{https://arxiv.org/abs/hep-ph/9608470}{{\tt hep-ph/9608470}}]

\bibitem{Kirschner:1983di}
R.~Kirschner and L.N.~Lipatov,
  \emph{{Double Logarithmic Asymptotics and Regge Singularities of Quark 
  Amplitudes with Flavor Exchange}},
  \href{https://dx.doi.org/10.1016/0550-3213(83)90178-5}
  {Nucl.\ Phys.\ B213 (1983) 122}

\bibitem{Bajnok:2008qj}
Z.~Bajnok, R.A.~Janik and T.~Lukowski,
  \emph{Four-loop twist two, BFKL, wrapping and strings},
  \href{https://doi.org/10.1016/j.nuclphysb.2009.02.005}
  {Nucl.~Phys.\ B816 (2009) 376}
  [\href{https://arxiv.org/abs/0811.4448}{{\tt arXiv:0811.4448}}]

\bibitem{vanNeerven:1991nn}
W.L.~van Neerven and E.B.~Zijlstra,
  \emph{{Order $\as(2)$ contributions to the deep inelastic Wilson 
  coefficient}},
  \href{https://doi.org/10..1016/0370-2693(91)91024-P}
  {Phys.\ Lett.\ B272 (1991) 127}

\bibitem{Zijlstra:1991qc}
E.B.~Zijlstra and W.L.~van~Neerven,
  \emph{{Contribution of the second order gluonic Wilson coefficient 
  to the deep inelastic structure function}},
  \href{http://dx.doi.org/10.1016/0370-2693(91)90301-6}
  {Phys.\ Lett.\ B273 (1991) 476}

\bibitem{Catani:1989sg}
S.~Catani, F.~Fiorani, G.~Marchesini,
 \emph{{Small-$x$ Behavior of Initial State Radiation in Perturbative QCD}},
 \href{https://dx.doi.org/10.1016/0550-3213(90)90342-B}
 {Nucl.\ Phys.\ B336 (1990) 18}

\bibitem{Catani:1994sq}
S.~Catani and F.~Hautmann, 
  \emph{{High-energy factorization and small-x deep
  inelastic scattering beyond leading order}},
  \href{http://dx.doi.org/10.1016/0550-3213(94)90636-X}
  {Nucl.\ Phys.\ B427 (1994) 475} 
  [\href{http://arxiv.org/abs/hep-ph/9405388}{{\tt hep-ph/9405388}}]

\bibitem{Fadin:1998py}
V.S.~Fadin and L.N.~Lipatov,
  \emph{{BFKL pomeron in the next-to-leading approximation}}, \\
  \href{https://dx.doi.org/10.1016/S0370-2693(98)00473-0}
  {Phys.\ Lett.\ B429 (1998) 127},
  [\href{https://arxiv.org/abs/hep-ph/9802290}{{\tt hep-ph/9802290}}]

\bibitem{Ciafaloni:1998gs}
M.~Ciafaloni and G.~Camici,
  \emph{{Energy scale(s) and next-to-leading BFKL equation}}, \\
  \href{https://dx.doi.org/10.1016/S0370-2693(98)00551-6}
  {Phys.\ Lett.\ B430 (1998) 349}
  [\href{https://arxiv.org/abs/hep-ph/9803389}{{\tt hep-ph/9803389}}]

\bibitem{Ciafaloni:2005cg}
M.~Ciafaloni and D.~Colferai,
  \emph{{Dimensional regularisation and factorisation schemes in the 
  BFKL equation at subleading level}},
  \href{https://doi.org/10.1088/1126-6708/2005/09/069}{JHEP 09 (2005) 069}
  [\href{https://arxiv.org/abs/hep-ph/0507106}{{\tt hep-ph/0507106}}

\bibitem{Ciafaloni:2006yk}
M.~Ciafaloni, D.~Colferai, G.P.~Salam, A.M.~Stasto,
   \emph{{Minimal subtraction vs.~physical factorisation schemes in 
   small-x QCD}},
   \href{https://doi.org/10.1016/j.physletb.2006.03.014}
   {Phys.\ Lett.\ B635 (2006) 320}
   [\href{https://arxiv.org/abs/hep-ph/0601200}{{\tt hep-ph/0601200}}]

\bibitem{Soar:2009yh}
G.~Soar, S.~Moch, J.A.M. Vermaseren and A.~Vogt, 
  \emph{{On Higgs-exchange DIS, physical evolution kernels and fourth-order 
  splitting functions at large x}}, 
  \href{http://dx.doi.org/10.1016/j.nuclphysb.2010.02.003}
  {Nucl.\ Phys.\ B832 (2010) 152}
  [\href{http://arxiv.org/abs/0912.0369}{{\tt arXiv:0912.0369}}]

\bibitem{Almasy:2010wn}
A.A.~Almasy, G.~Soar and A.~Vogt,
 \emph{{Generalized double-logarithmic large-x resummation in inclusive 
 deep-inelastic scattering}},
 \href{https://doi.org/10.1007/JHEP03(2011)030}{JHEP 03 (2011) 030}
 [\href{https://arxiv.org/abs/1012.3352}{{\tt arXiv:1012.3352}}]

\bibitem{Vogt:2011jv}
A.~Vogt,
  \emph{{Resummation of small-x double logarithms in QCD: 
  semi-inclusive electron-positron annihilation}},
  \href{https://doi.org/10.1007/JHEP10(2011)025}
  {JHEP 10 (2011) 025}
  [\href{https://arxiv.org/abs/1108.2993}{{\tt arXiv:1108.2993}}]

\end{thebibliography}
\end{document}